\documentclass[twocolumn,english,aps,prd,10pt,floatfix]{revtex4-1}
\usepackage[T1]{fontenc}
\usepackage[latin9]{inputenc}
\usepackage{xcolor}
\usepackage{amsmath}
\allowdisplaybreaks
\usepackage{graphicx}
\usepackage{amssymb}
\usepackage{slashed}
\usepackage{hyperref}
\usepackage{dsfont}
\usepackage{enumerate}

\hypersetup{
    colorlinks,
    linkcolor={blue},
    citecolor={blue},
    urlcolor={blue}
}

\graphicspath{{.}{../}{figs/}{../figs/}}

\newcommand{\rme}{\mathrm{e}}
\newcommand{\rmi}{\mathrm{i}}
\newcommand{\rmd}{\mathrm{d}}

\newcommand{\NQ}{N}

\newcommand{\Nf}{N_{\text{f}}}

\renewcommand{\vec}[1]{\boldsymbol{#1}}

\newcommand{\sr}[1]{{\color{black}#1}}
\newcommand{\lj}[1]{{\color{black}{#1}}}

\begin{document}

\title{
Quantum critical behavior of two-dimensional Fermi systems with quadratic band touching
}

\author{Shouryya Ray}
\author{Matthias Vojta}
\author{Lukas Janssen}
\affiliation{Institut f\"ur Theoretische Physik, Technische Universit\"at Dresden,
01062 Dresden, Germany}

\date{19 December 2018}

\begin{abstract}
We consider two-dimensional Fermi systems with quadratic band touching and $C_3$ symmetry, as realizable in Bernal-stacked honeycomb bilayers. 
Within a renormalization-group analysis, we demonstrate the existence of a quantum critical point at a finite value of the density-density interactions, separating a semimetallic disordered phase at weak coupling from a gapped ordered phase at strong coupling.
The latter may be characterized by, for instance, antiferromagnetic, quantum anomalous Hall, or charge density wave order.
In the semimetallic phase, each point of quadratic band touching splits into four Dirac cones as a consequence of the nontrivial interaction-induced self-energy correction, which we compute to the two-loop order. 
We show that the quantum critical point is in the $(2+1)$-dimensional Gross-Neveu universality class characterized by emergent Lorentz invariance and a dynamic critical exponent $z=1$.
At finite temperatures, $T > 0$, we hence conjecture a crossover between $z=2$ at intermediate~$T$ and $z=1$ at low~$T$, and we construct the resulting nontrivial phase diagram as function of coupling strength and temperature.
\end{abstract}


\maketitle


\section{Introduction}
\label{sec:intro}

Interacting electron systems whose Fermi surfaces comprise isolated points in momentum space have proven to be a fertile subject of study, being host to a fascinating interplay of band topology and interactions. Interest in such Fermi-point systems was ignited by the discovery of massless Dirac quasiparticles in monolayer graphene~\cite{castrorev}. Upon the inclusion of weak short-range interactions, the Dirac semimetal state is stable.  At strong coupling, on the other hand, the system undergoes a quantum phase transition of the $(2+1)$-dimensional Gross-Neveu universality class towards a massive-fermion phase, characterized, for instance, by antiferromagnetic, quantum anomalous Hall, or charge density wave order, depending on the microscopic character of the interactions~\cite{herbut06, herbut09, assaad13, janssen14, otsuka16}.
Fermi-point systems with quadratic band touching (QBT) have begun to garner much attention as well~\cite{ref6, herbut14}. A paradigmatic example is given by the nearest-neighbor-hopping model on the Bernal-stacked bilayer honeycomb lattice, a simple model for bilayer graphene \cite{McCannFalko}. The enhanced density of states as compared to Dirac-point systems makes these systems more susceptible to the effects of interactions.
In particular, in two spatial dimensions, repulsive short-range interactions are marginally relevant in the renormalization-group (RG) sense, implying a runaway flow and spontaneous symmetry breaking already at infinitesimal values of the microscopic couplings \cite{ref6}.
An extended stable semimetallic phase, with a nontrivial quantum phase transition towards a symmetry-broken state at a finite value of the coupling, has therefore long been thought to be impossible in systems with QBT.
However, unlike in checkerboard and Kagome lattices, the point of QBT in bilayer honeycomb systems is not protected by symmetry. This is because the honeycomb lattice has only a $C_3$ rotational symmetry, which allows in principle a splitting of the QBT point into four Dirac cones~\cite{ref6}. In fact, this is precisely what happens when interlayer hopping terms beyond the shortest range are taken into account (so-called trigonal warping terms)~\cite{McCannFalko}.

\begin{figure}
\includegraphics[width=\linewidth]{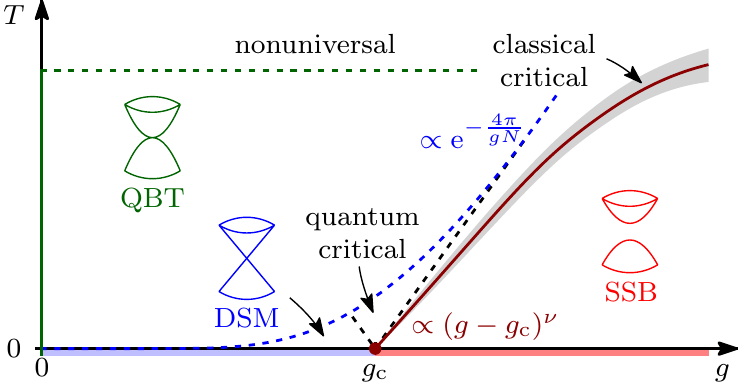}
\caption{Schematic phase diagram of QBT systems with $C_3$ symmetry as a function of temperature $T$ and short-range interaction $g$. Dashed curves denote crossovers, the solid curve denotes the finite-temperature phase transition for the case of discrete symmetry breaking. The dashed green horizontal line separates the universal regime at intermediate and low temperatures from the nonuniversal high-temperature regime. At intermediate temperatures, the fermion spectrum is effectively quadratic, characterized by dynamic exponent $z = 2$ (QBT). At temperatures below the blue dashed curve, the flow enters the Dirac regime with $z=1$ (DSM). The black dashed lines emerging from the critical point at $g_\mathrm{c}$ denote the quantum critical regime, characterized by a continuum of excitations. The transition towards the ordered phase occurs at finite temperature in the case of discrete spontaneous symmetry breaking (SSB). The critical temperature $T_\mathrm{c} \propto (g-g_\mathrm{c})^\nu$ is shown as solid red curve, together with its concomitant classical critical regime (gray shaded). Details are discussed in Sec.~\ref{sec:discussion}.}
\label{fig:finite-T}
\end{figure}

In this work, we demonstrate explicitly that even in the case when the trigonal warping terms are absent in the microscopic Hamiltonian, the presence of higher-order terms, despite being irrelevant in the RG sense, generate effective trigonal warping terms at low energy. This leads to a stable semimetallic phase at weak short-range interactions and a nontrivial quantum critical point at finite coupling. We establish the relevant $(2+1)$-dimensional Gross-Neveu universality class for the transition and map out the pertinent phase diagram in the plane of temperature $T$ and interaction strength $g$. This is depicted in Fig.~\ref{fig:finite-T}.
Simple estimates for the size of the interactions place suspended bilayer graphene on the ordered side of the transition. At intermediate temperatures, the two points of QBT in bernal-stacked bilayer graphene should split into eight Dirac cones before the low-temperature instability develops. The nature of the ordered ground state sensitively depends on the relative size of the various possible interactions~\cite{Cvetkovic}. Candidates for the low-temperature state include antiferromagnets, quantum anomalous Hall states, nematic states, and charge density waves~\cite{rozhkov16}. Most experiments point to an insulating state with a full gap in the spectrum~\cite{feldman09, martin10, weitz10, velasco12, freitag12, bao12, veligura12} (see, however, Ref.~\cite{mayorov11}).
Our work suggests that bilayer graphene may show vestiges of the quantum critical scaling in the regime above the transition temperature $T_\mathrm{c}$. These should be observable in various transport quantities, such as the Hall coefficient $R_\mathrm{H}(T)$. We predict a crossover from $R_\mathrm{H} \propto T^{-1}$ in the QBT regime for $T \gg T_\mathrm{c}$ to $R_\mathrm{H} \propto T^{-2}$ in the Dirac regime for $T \gtrsim T_\mathrm{c}$, before eventually the gap opens up and $R_\mathrm{H}$ will follow an exponential law for $T < T_\mathrm{c}$.

Our result confirms the schematic RG picture previously purported by Pujari \textit{et al.}~\cite{Pujarimain} in the context of quantum Monte Carlo simulations of the Hubbard model on the bilayer honeycomb lattice. 
In the absence of trigonal warping, the numerics pointed to an extended gapless phase at weak coupling and a quantum critical point to a gapped ordered phase at a finite Hubbard interaction. The measured values for the dynamic critical exponent $z = 0.9(2)$ and the correlation-length exponent $\nu = 1.0(2)$ were broadly consistent with the $(2+1)$-dimensional Gross-Neveu universality class, the particular type of which, however, had not been possible to establish unambiguously.
As proposed already in Ref.~\cite{Pujarimain} (see also Ref.~\cite{honerkamp17b}), the crucial ingredients for this mechanism are the interaction-induced corrections to the fermion self-energy.
However, at the one-loop order, which has been thoroughly investigated in previous works~\cite{ref6, zhang10, vafekyang, vafekbig, uebelacker11, lang12, scherer12, Cvetkovic, song12}, the self-energy correction happens to vanish as a consequence of the interaction being local. (The one-loop correction is finite, however, if long-range interactions are present \cite{sinnerziegler}). A field-theoretic understanding of the quantum critical behavior seen in the numerics therefore requires to go beyond the one-loop order, which is a daunting task due to the absence of relativistic and continuous rotational symmetries. As a result, a proper RG analysis of this physics has, to the best of our knowledge, thus far been lacking in the literature. It is one of our main technical advances to demonstrate that the two-loop self-energy corrections can be computed in an analytical way by employing a suitably adapted regularization scheme in \emph{position} space.
We construct a minimal continuum low-energy field-theory that captures the salient physics of interacting $C_3$ symmetric QBTs. We then evaluate all loop corrections to the leading non-vanishing order. This, most importantly, includes the crucial two-loop self-energy diagrams and it allows us to derive improved RG flow equations. This leads us to construct the corresponding quantum phase diagram and to reveal the pertinent universality class and its critical exponents. We also compare with mean-field solutions, which are controlled in a certain large-$N$ limit, and discuss the behavior at finite trigonal warping on the microscopic level.

The body of this paper is organized as follows: Section~\ref{sec:lattice} introduces the minimal effective field theory starting from the tight-binding model on a Bernal-stacked bilayer honeycomb lattice. Mean-field solutions are studied in Sec.~\ref{sec:mean-field}. In Sec.~\ref{sec:rg}, we then proceed to evaluating the leading loop corrections and investigate the phase diagram arising from the RG flow equations. 
Critical exponents and the finite-temperature phase diagram are discussed in Sec.~\ref{sec:discussion}.
The paper closes with a conclusion and an outlook in Sec.~\ref{sec:concl}. Technical details are deferred to two appendices.


\section{From lattice to low-energy field theory}
\label{sec:lattice}
In this section, we wish to motivate a minimal continuum field theory that shall be the main object of study in this paper. For concreteness, we start with a specific microscopic model on a lattice with $C_3$ symmetry and derive thence a Euclidean action serving as an effective low-energy description. The pure QBT theory with $z=2$ on the one hand and the relativistic Gross-Neveu theory with $z=1$ on the other hand are recovered from this continuum field theory in two opposite limiting cases.
We would like to emphasize, however, that the physics we are investigating is independent of the particular lattice model and quite generally applies to any interacting two-dimensional Fermi system with QBT and $C_3$ rotational symmetry.


\subsection{Fermions on Bernal-stacked honeycomb bilayer}
\label{sec:tightbinding}

Consider a model of spinless fermions on the Bernal-stacked bilayer honeycomb lattice at half filling, defined by the tight-binding Hamiltonian~\cite{castrorev}
\begin{align}
	H_0 &= -t\sum_{\langle ij \rangle}\sum_{m=1}^2 a_{im}^\dagger b_{jm} - t_\perp \sum_{i} a_{i1}^\dagger b_{i2}
	\nonumber \\ & \qquad
	-t_{\text{w}} \sum_{\langle ij \rangle} a_{i1}^\dagger b_{j2} + \textnormal{H.c.}
	\label{eq:tight-binding-realspace}
\end{align}
The operators $a_{im}$ ($b_{im}$) annihilate a fermion in layer $m$ and sublattice A (B) at position $\vec{R}_i$ of the Bravais lattice. 
The parameter $t$ corresponds to hopping processes between nearest neighbors $\langle i j\rangle$ within the same honeycomb layer, while $t_\perp$ corresponds to hopping between sites that are located on top of each other and belong to different layers and different sublattices. The third term in $H_0$ parametrized by $t_\text{w}$ denotes the trigonal warping term \sr{allowed by $C_3$ symmmetry} and corresponds to next-nearest-neighbor interlayer hopping processes.
We denote the primitive Bravais lattice vectors as $\vec{a}_1 = (1/2,\sqrt{3}/2)$ and $\vec{a}_2 = (1/2,-\sqrt{3}/2)$, where we have set the lattice constant $a = 1$ for notational simplicity. Proper units of $a$ will be restored below whenever needed. In reciprocal space and upon collecting the Fourier-transformed fermion operators into a vector $c^\dagger(\vec{k}) = \left(a_1^\dagger(\vec{k}),b_2^\dagger(\vec{k}),a_2^\dagger(\vec{k}),b_1^\dagger(\vec{k})\right)$, the tight-binding Hamiltonian can be written in matrix notation as
\begin{align}
	H_0 &= \int_{\vec k \in \text{BZ}}\frac{\rmd^2 \vec{k}}{(2\pi)^2}\,c^\dagger(\vec{k}) \, \mathcal{H}_0(\vec{k}) \, c(\vec{k}),
\end{align}
where the $\vec k$-integration is over the Brillouin zone (BZ). The Hermitian $4 \times 4$ matrix $\mathcal H_0$ reads in block notation
\begin{align}
  \mathcal{H}_0(\vec{k}) &=
    \begin{pmatrix}
     \mathcal{H}_{11}(\vec{k}) & \mathcal{H}_{12}(\vec{k}) \\
     \mathcal{H}_{12}^\dagger(\vec{k}) & \mathcal{H}_{22}(\vec{k})
    \end{pmatrix},
\end{align}
with the $2\times 2$ blocks having nonvanishing entries only on the off-diagonal,
\begin{align}
  \mathcal{H}_{11}(\vec{k}) &= 
  	-t_{\text{w}} \begin{pmatrix}
	0 & f^*(\vec{k}) \\
  	f(\vec{k}) & 0
  	\end{pmatrix},\\
  \mathcal{H}_{12}(\vec{k}) &= 
	-t \begin{pmatrix}
		0 & f(\vec{k}) \\
		f^*(\vec{k}) & 0
	\end{pmatrix}, \\
  \mathcal{H}_{22}(\vec{k}) &= 
	-t_\perp \begin{pmatrix}
	0 & 1 \\
	1 & 0
	\end{pmatrix}.
\end{align}
Here, $f(\vec{k}) = \sum_{\boldsymbol{\delta}} \rme^{\rmi \vec{k}\cdot \boldsymbol{\delta}}$ is the nearest-neighbor form factor of the honeycomb lattice, with $\boldsymbol{\delta} \in \{(1,0),\vec{a}_1,\vec{a}_2\}$ the three nearest-neighbor displacement vectors. The spectrum of $\mathcal{H}_0(\vec k)$ consists of four bands with dispersion $\pm \varepsilon_\pm(\vec k)$, given by
\begin{align}
  \varepsilon_\pm^2(\vec k) & = \frac{1}{2}\Big[ t_\perp^2 + (2t^2 + t_{\text{w}}^2) |f|^2 \pm \bigl\{t_\perp^4 + t_{\text{w}}^2\left(4 t^2 + t_{\text{w}}^2\right)|f|^4 
  \nonumber \\ & \quad
  + 2t_\perp |f|^2 \left(2 t^2 t_\perp - t_\perp t_{\text{w}}^2 + 4 t^2 t_{\text{w}} \operatorname{Re}f \right) \bigr\}^{\!1/2}\Big].
  \label{eq:dispfull}
\end{align}
Here, we have suppressed the momentum dependence of the form factor $f \equiv f(\vec k)$ for notational brevity. \sr{The above spectrum exhibits particle-hole symmetry, which we shall assume throughout from hereon in. We note that it \lj{will} be broken upon inclusion of longer-ranged terms in the tight-binding Hamiltonian \eqref{eq:tight-binding-realspace}, \lj{such as next-nearest neighbor intralayer hopping \cite{McCannKoshino}}. The additional physics due to broken particle-hole symmetry \lj{is interesting in its own right and will be left for future work}.} %

At half filling and low temperatures, only the two bands at $\pm \varepsilon_-(\vec k)$ contribute to physical observables. The general properties of the spectrum now depend crucially on whether the trigonal warping $t_{\text{w}}$ is finite or vanishes, so we discuss these two cases separately in the following.

\begin{figure*}
\includegraphics[width=\textwidth]{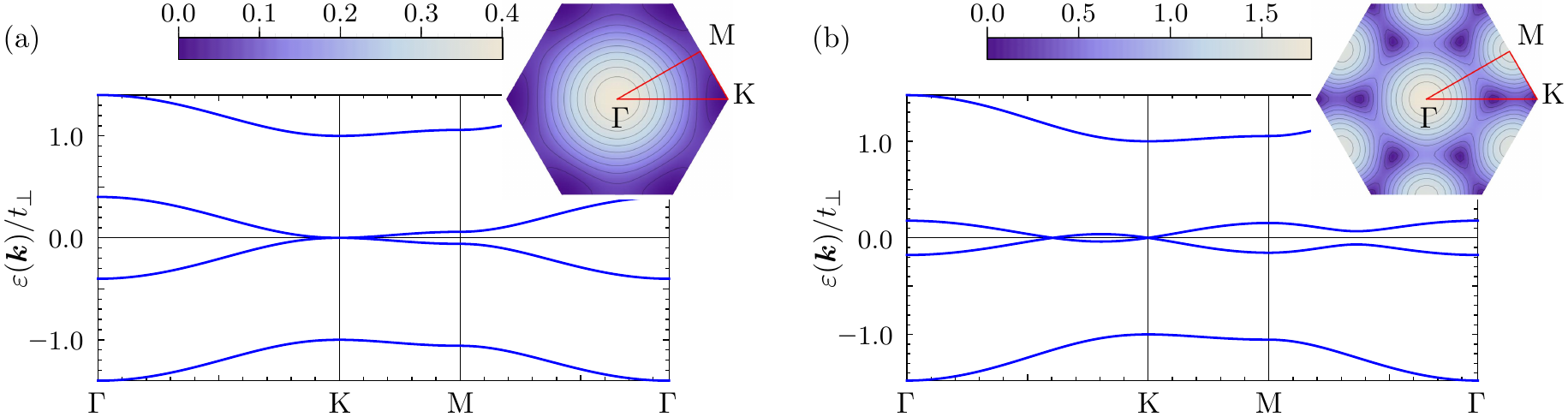}
\caption{Tight-binding dispersion along the high-symmetry line $\Gamma$--$K$--$M$--$\Gamma$ from Eq.~\eqref{eq:dispfull} for $t/t_\perp$ = 0.25. The insets show the dispersion of the low-energy conduction band in the first Brillouin zone (color plot) and the path used in the main panels (red line). 
In (a), there is no trigonal warping, $t_{\text{w}}/t_\perp = 0$, and only the two QBTs at $\vec{k} = \pm\vec{K}$ touch the Fermi level at $\varepsilon = 0$.
For nonzero $t_{\text{w}}/t_\perp = 0.1$ (b), the two QBTs split into two central Dirac cones at $\vec{k} = \pm\vec{K}$ and six ``satellite'' Dirac cones at incommensurate wavevectors between $\Gamma$ and $K$.}
\label{fig:lattice_disp}
\end{figure*}

First, when the trigonal warping is tuned to zero, $t_{\text{w}} \to 0$, the two low-energy bands touch at $\vec{k} = \pm \vec{K}$, where $\vec{K} = (4\pi/3,0)$ denotes the high-symmetry $K$ point at one of the corners of the hexagonal Brillouin zone. To see that these two band crossings are indeed quadratic, we expand the form factor around $\pm\vec{K}$ as
\begin{align}
	f(\pm\vec{K} + \vec{p}) &= \mp \frac{\sqrt{3}}{2} |\vec p| \operatorname{e}^{\mp \rmi \varphi} + \frac{1}{8} |\vec p|^2 \operatorname{e}^{\pm 2\rmi \varphi} + \mathcal O(|\vec p|^3),
\end{align}
where $\varphi = \arg(p_x + \rmi p_y)$ denotes the polar angle of the local momentum $\vec p  = \vec k \mp \vec K$.
Upon subsequent expansion of the low-energy spectrum to next-to-leading order in $|\vec p|$, one finds
\begin{align}
	\varepsilon^2_-(\pm \vec K + \vec{p}) = \frac{9t^4}{16t_\perp^2}|\vec{p}|^4\left(1 \mp \frac{|\vec{p}|}{\sqrt{3}} \cos 3\varphi \right) + \mathcal O(|\vec p|^6),
\label{eq:dispQBT}
\end{align}
valid for $|\vec p| \ll t_\perp/t$.
This demonstrates the existence of two twofold degenerate QBT points located at the two inequivalent $K$ points in the Brillouin zone and at energy $\varepsilon_-(\pm \vec K) = 0$. By inspection of the full band structure given by Eq.~\eqref{eq:dispfull}, one readily finds that there are no further bands crossing the zero-energy level, see Fig.~\ref{fig:lattice_disp}(a). In the half-filled case, the Fermi level is therefore fixed precisely at the two QBT points.
We note that the leading-order term $\propto |\vec p|^4$ in Eq.~\eqref{eq:dispQBT} exhibits a continuous $\mathrm O(2)$ rotational symmetry in momentum space. The next-to-leading order term $\propto |\vec p|^5 \cos 3\varphi$, by contrast, breaks this symmetry explicitly down to $C_3$, reflecting the lattice symmetry of the honeycomb model.
At the level of the free theory, the $\mathrm O(2)$ rotational symmetry therefore emerges dynamically if one restricts the window of observation to sufficiently low energies.
However, we shall demonstrate that this is no longer true once interactions are taken into account.

We now switch on a small finite $t_{\text{w}} > 0$. We again expand the spectrum to next-to-leading order in local momentum, but now we keep the leading $t_{\text{w}}$ correction in each power of $|\vec{p}|$. The low-energy spectrum then takes the form
\begin{align}
  \varepsilon^2_-(\pm \vec K + \vec{p}) = \frac{3t_{\text{w}}^2}{4}|\vec{p}|^2\left(1 \pm \frac{\sqrt{3}t^2}{t_\perp t_{\text{w}}}|\vec{p}|\cos 3\varphi \right) + \mathcal O(|\vec p|^4).\label{eq:varepsilonDir}
\end{align}
Note the lower exponent of the leading-order term as compared to Eq.~\eqref{eq:dispQBT}.
Consequently, the local dispersion near $\pm\vec K$ is no longer quadratic, but linear, and the spectrum exhibits Dirac cones at $\vec k = \pm \vec K$. In addition, for each Dirac cone at one of the high-symmetry $K$ points, there are three ``satellite'' Dirac cones located at incommensurable wavevectors $\vec k = \pm \vec K + \vec p$ with
\begin{align}
|\vec{p}| = \frac{4t_\perp t_{\text{w}}}{\sqrt{3} t^2} \quad \text{and} \quad \varphi = (4n+1 \pm 1)\frac{\pi}{6},
\label{eq:locDirac}
\end{align}
where $n = 0,1,2$ and we have assumed $t_\perp > 0$ and $t_\text{w} > 0$ for concreteness.
The full dispersion in the presence of trigonal warping is illustrated in Fig.~\ref{fig:lattice_disp}(b). %
The splitting of each QBT point into four Dirac cones upon switching on a small finite $t_\text{w}$ can be understood on more general grounds~\cite{ref6}: A QBT carries Berry flux $\pm2\pi$, while a Dirac cone has Berry flux $\pm\pi$. Let us take the QBT at $\vec k = +\vec{K}$, whose Berry flux is $+2\pi$. Due to flux conservation, the QBT can be either split into two Dirac cones carrying $+\pi$ flux, or one $-\pi$ Dirac cone and three $+\pi$ Dirac cones (as long as no more Dirac cones are involved).
On the honeycomb lattice, the realization that is compatible with the $C_3$ symmetry is the splitting of the QBT at $\vec k = + \vec K$ into one $- \pi$ Dirac cone at $\vec k = +\vec K$ and three Dirac cones shifted along the lines through $K$ and the centers $\Gamma$ of the three neighboring Brillouin zones, in agreement with Eq.~\eqref{eq:locDirac}. %
Hence, the total number of Dirac cones per QBT and their location in the Brillouin zone is a consequence of the conservation of Berry flux combined with the symmetry of the honeycomb lattice. \sr{Finally, the fact that the satellite Dirac cones lie at the same energy as the ones at the $K$-points is a consequence of particle-hole symmetry.}%

This concludes the discussion at the non-interacting level. In particular, the fermiologies in the QBT and Dirac cases are distinct, and going from the former to the latter requires ``switching on'' a parameter like $t_{\text{w}}$ by hand, as done above. In the presence of fermion-fermion interactions, however, this occurs dynamically. To elucidate this further, we now construct a pertinent low-energy continuum field theory.


\subsection{Continuum limit}
\label{sec:action}

For the non-interacting part, we begin by writing down the Hamiltonian for QBT in the case of $t_\text{w} = 0$. In a $4 \times 4$ representation, it can be written as~\cite{ref6, vafekbig, janssenherbutQBT}
\begin{align}
\mathcal{H}^{(2)}_0(\vec{p}) = d_a(\vec{p}) \, (\sigma^a \otimes \mathds 1_2), \qquad a = 1,2,
\label{eq:QBTminimal}
\end{align}
where we have assumed the summation convention over repeated indices. In the above equation, the diagonal factor $\mathds 1_2$ can be understood to act on the valley index.
The $2 \times 2$ matrices $\sigma^a$ anticommute with each other and square to one, and may be represented by the usual Pauli matrices, $\sigma^1 \equiv \sigma_x$ and $\sigma^2 \equiv \sigma_y$.
The time-reversal operator can then be defined as $T = (\sigma_x \otimes \sigma_x) \mathcal K$, where $\mathcal K$ denotes complex conjugation.
The functions $d_a(\vec{p})$ are $\vec{p}^2$ times the real spherical harmonics of angular momentum $\ell = 2$, which in two dimensions simply become $d_1(\vec{p}) = p_x^2 - p_y^2 = \vec p^2 \cos 2\varphi$ and $d_2(\vec p) = 2p_x p_y = \vec p^2 \sin 2\varphi$. 

Under $\operatorname{O}(2)$ spatial rotations with angle $\theta$,
\begin{align}
 p^a \mapsto (R_\theta)^{a}{}_{b}\,p^b,
 \quad
 R_\theta = 
 \begin{pmatrix}
 \cos \theta & -\sin \theta \\
 \sin \theta & \cos \theta
 \end{pmatrix} \in \mathrm{O}(2),
\end{align}
the $d_a$ and $\sigma^a$ transform respectively as
\begin{align}
d_a(\vec{p}) &\mapsto (R_{2\theta})_a^{\phantom{a}b}\,d_b(\vec{p}), &
\sigma^a &\mapsto (R_{2\theta})^{a}_{\phantom{a}b}\,\sigma^{b}.
\end{align}
While the former equation follows from direct computation, the latter is to be understood in the sense that the $\sigma^a$ transform as components of the second-rank tensor~\cite{janssenherbutQBT}
\begin{align}
  \begin{pmatrix}
    \sigma^1 & \sigma^2 \\
    \sigma^2 & -\sigma^1
  \end{pmatrix}
  \mapsto R_\theta^\top 
  \begin{pmatrix}
    \sigma^1 & \sigma^2 \\
    \sigma^2 & -\sigma^1
  \end{pmatrix}
  R_\theta.
\label{eq:tensorranktwo}
\end{align}
This shows that $\mathcal H_0^{(2)}$ is invariant under $\operatorname{O}(2)$ rotations.
With the above definitions, it is straightforward to verify that the product $\sigma^1 \sigma^2$ is also invariant under rotations. Consequently, the two remaining matrices $\sigma^0 \equiv \mathds{1}_2$ and $\sigma^3 = -\rmi \sigma^1 \sigma^2$ that together with $\sigma^1$ and $\sigma^2$ span the space of $2\times2$ matrices, are rotationally invariant. At the quadratic order $\mathcal O(|\vec p|^2)$, therefore, the only possible term in the Hamiltonian that is compatible with the $C_3$ symmetry and diagonal in valley space is the $\mathrm O(2)$ invariant one present in the above $\mathcal H_0^{(2)}$. 
The upshot is that any free 2D Fermi system with QBT and $C_n$ symmetry with $n \geqslant 3$ has emergent $\operatorname{O}(2)$ symmetry at low energies.

At the linear order $\mathcal O(|\vec p|)$, however, a $C_3$ invariant term that breaks $\mathrm{O}(2)$ is perfectly possible. For instance, the term
\begin{align} \label{eq:H0-1}
  \mathcal{H}_0^{(1)}(\vec{p}) = \overline{p}_a (\sigma^a \otimes \sigma^3)
\end{align}
with $\overline{\vec{p}} \equiv (\overline{p}^a) = (p_x, -p_y)^\top$ transforms under rotations as $\mathcal H_0^{(1)}(\vec p) \mapsto \overline p_a (R_{3\theta})^a{}_b \sigma^b$ and is therefore only symmetric under the $C_3$ symmetry, but not continuous $\mathrm{O}(2)$ rotations.

At the cubic order $\mathcal O(|\vec p|^3)$, an analogous term is $C_3$ symmetry allowed,
\begin{align} \label{eq:H0-3}
  \mathcal{H}_0^{(3)}(\vec{p}) = \vec{p}^2 \, \overline{p}_a (\sigma^a \otimes \sigma^3),
\end{align}
which manifestly has the same symmetry properties as $\mathcal H_0^{(1)}$.

A general noninteracting low-energy Hamiltonian consistent with $C_3$ rotational symmetry can therefore be written in terms of three parameters $f_1$, $f_2$, and $f_3$ as
\begin{align}
	\mathcal{H}_0(\vec{p}) & = 
	\sigma^a \otimes \left[
	f_1 \overline p_a \sigma^3
	+ f_2 d_a(\vec p) \mathds 1_2
	- f_3 \vec p^2 \overline p_a \sigma^3
	\right] 
	\nonumber \\ & \quad
	+ \mathcal O(|\vec p|^4),
	\label{eq:fullH0}
\end{align}
where the signs of $f_1,f_2$ and $f_3$ have been chosen for later convenience.
The spectrum of $\mathcal H_0(\vec p)$ is given by
\begin{align}
	\varepsilon^2(\vec p) & = f_1^2 |\vec p|^2 + 2f_1 f_2 |\vec p|^3 \cos 3\varphi + (f_2^2 -  2 f_1 f_3) |\vec p|^4 
	\nonumber \\ & \quad
	- 2 f_2 f_3 |\vec p|^5 \cos 3\varphi + \mathcal O(|\vec p|^6),
\end{align}
and hence reproduces the tight-binding dispersion near the $K$ point at $\vec k = + \vec K$ [Eqs.~\eqref{eq:dispQBT} and \eqref{eq:varepsilonDir}] for 
\begin{align}
	f_1 & = \frac{\sqrt{3} t_\text{w}a}{2}, &
	f_2 & = \frac{3 t^2 a^2}{4 t_\perp}, &
	f_3 & = \frac{a^3}{2\sqrt{3}} \frac{3 t^2}{4 t_\perp},
	 \label{eq:fm10fromTB}
\end{align}
and the same equations hold, up to a suitable change of the local momentum basis $\vec p \mapsto \overline{\vec p}$, near the second $K$ point at $\vec k = - \vec K$ as well.
Here, we have reinstated the lattice constant $a$ in order to make the physical units more readily apparent.
In the following, we shall in particular be interested in the situation in which $f_1$ is tuned to zero at the microscopic level (which corresponds to $t_\text{w} = 0$ in the tight-binding Hamiltonian) describing a system \sr{whose bare spectrum has a} QBT \sr{(referred to henceforth as ``the QBT limit'')}, and study the dynamical generation of $f_1$ due to interactions.

The Lagrangian is constructed from Eq.~\eqref{eq:fullH0} in canonical fashion, namely
\begin{align}
\mathcal{L}_0 = \psi^\dagger_i \left[\partial_\tau + \mathcal{H}_0(-\rmi \nabla)\right]\psi^i,
\label{eq:L0}
\end{align}
where $\tau$ denotes imaginary time and $\psi^i$, $\psi^\dagger_i$ are four-component complex spinors with ``flavor'' index $i = 1,\dots,\Nf$. On the honeycomb bilayer and in the limit of vanishing trigonal warping $t_\text{w} \ll t^2 / t_\perp$, for which the spectrum has a QBT, the flavor number $\Nf$ can be understood as the real-spin degeneracy of each band. Thence, $\Nf = 1$ for spinless fermions. For the sake of generality, however, we shall keep the flavor number $\Nf$ arbitrary in our calculations. This also allows us to make contact with the limiting cases $\Nf \to \infty$, which represent the mean-field limit, and $\Nf = 1/2$, which can be understood as a Fermi system with a single point of QBT in the Brillouin zone, as realizable for spinless fermions on the Kagome and checkerboard lattices \cite{ref6}.
Note, however, that for the latter systems, the linear and cubic terms in Eqs.~\eqref{eq:H0-1} and \eqref{eq:H0-3} are forbidden by time-reversal symmetry and the QBT is therefore protected for $\Nf = 1/2$.

We emphasize that the above Hamiltonian $\mathcal{H}_0$, with the correct interpretation, is sufficient to capture the behavior at substantial trigonal warping as well. 
In this limit, Eq.~\eqref{eq:fullH0} describes massless Dirac fermions subject to a quadratic perturbation $\propto f_2$, with the spectrum given by Eq.~\eqref{eq:varepsilonDir}. Some care is needed when it comes to the flavor content of the low-energy Dirac theory. Since a separate fermion flavor has to be introduced for each Fermi point, one requires four Dirac points for every valley in the QBT theory. Flavor symmetry between the ``satellite'' and the central Dirac point can be restored by a suitable rescaling of the local momentum, viable in the low-energy limit.

In conclusion, therefore, the Lagrangian \eqref{eq:L0} constitutes two different continuum field theories describing two opposite limits of the low-energy physics of fermions on the bilayer honeycomb lattice: On the one hand, the QBT limit for vanishing or infinitesimally small trigonal warping $t_\text{w} \ll t^2/t_\perp$ is described by Eq.~\eqref{eq:L0} with flavor number $\Nf = N/2$, where $N$ is the number of valleys in the QBT limit. On the other hand, the Dirac limit for $t_\text{w} \gg t^2/t_\perp$ is described by the same Eq.~\eqref{eq:L0} in the low-energy limit with, however, now $\Nf = 2N$ (Dirac) fermion flavors.
Hence, the number of four-component fermion flavors in the low-energy description is
\begin{align}
	\Nf = 
	\begin{cases}
		N/2 & \text{for } f_1/f_2 \ll 1/a, \\
		2N & \text{for } f_1/f_2 \gg 1/a.
	\end{cases}
\end{align}
As noted above, the concrete lattice realization of spinless fermions on a honeycomb bilayer corresponds to $N = 2$.


\subsection{Interactions}

A generic four-fermion interaction can be written in the form
\begin{align}
	\tfrac{1}{2}g_{rs}(\psi^\dagger_i A_r^{ij} \psi_j)(\psi^\dagger_k A_s^{kl} \psi_l)
\end{align}
with coupling parameters $g_{rs}$, where $r,s = 1,\ldots,16$~\cite{herbut09a, gies10}. The smallest subspace closed under the RG flow for $N \geqslant 2$ consists of three independent (i.e., Fierz irreducible) couplings \cite{vafekbig}. The nature of the concrete state that emerges upon spontaneous symmetry breaking is sensitive to the form of interactions present microscopically in the system. Our primary interest, however, is in the question whether spontaneous symmetry breaking takes place at all for small couplings, rather than the competition (or cooperation) between the different possible orders. For this purpose, it is sufficient to restrict ourselves to a single interaction channel. For definiteness, we choose $A_{r}^{ij} = A_{s}^{ij} = (\sigma^3 \otimes \sigma^3) \delta^{ij}$, corresponding to
\begin{align}
  \mathcal{L}_{\text{int}} = -\tfrac12 g\!\left[\psi^\dagger_i\!\left(\sigma^3 \otimes \sigma^3\right)\!\psi^i\right]^{\!2},
\end{align}
where we have adopted a sign convention that leads to a stabilization of the ordered state for positive values of~$g$. 
This particular choice of $\mathcal{L}_{\text{int}}$ is natural and appropriate for the following reasons: 
Firstly, note that $\sigma^3 \otimes \sigma^3$ anticommutes with $\mathcal H_0$. A finite bilinear condensate in the above interaction channel, i.e., $\langle \psi^\dagger_i (\sigma^3 \otimes \sigma^3) \psi^i \rangle \neq 0$, would therefore correspond to a state with a full mass gap in the spectrum, which is typically energetically favored within mean-field treatments~\cite{ref6, herbut14, janssenherbutQBT}.
The state is characterized by an imbalance of the number of particles on layer 1 compared to layer 2 and thus breaks inversion symmetry between the layers. Time reversal, by contrast, remains intact and the new ground state thus represents a topologically trivial interaction-induced insulator.
In fact, precisely this interaction channel has been found as the dominant ordering tendency in the $t$-$V$ model of spinless fermions on the Bernal-stacked honeycomb bilayer subject to a repulsive nearest-neighbor interaction $V$ within a multi-channel RG analysis~\cite{vafekbig}.
Secondly, this channel is readily identified with the simplest possible Lorentz scalar, $[\psi^\dagger_i (\sigma^3 \otimes \sigma^3) \psi^i]^2 \equiv (\overline{\psi}_i\psi^i)^2$, where $\overline{\psi}_i = \psi^\dagger_i (\sigma^3 \otimes \sigma^3)$ is the Dirac conjugate. This is the familiar Gross-Neveu-Ising interaction, which in the Dirac limit $t_\text{w} \gg t^2/t_\perp$ has a well-understood quantum critical point at finite $g$~\cite{hands93, vasiliev93, gracey94, vojta00a, vojta00b, braun11, gracey16, mihaila17, zerf17, iliesiu18, ihrig18}.
Furthermore, to leading order in $1/N$, it turns out that the above interaction channel is closed under RG in the sense that no further interactions are generated upon integrating out high-energy modes if absent on the microscopic level. We also note that for $N = 1$, in which case there are in total only two spinor components in the QBT limit, any finite four-fermion interaction must be proportional to $\psi^\dagger \sigma^3 \psi$. The single-channel approximation is therefore exact not only for $N \to \infty$, but also at $N = 1$. Although bilayer graphene with $N=2$ falls in neither class, we expect our major conclusions, concerning in particular the existence of a quantum critical point at finite coupling in the QBT limit, to hold also in this case. 
We shall briefly comment on the effect of other interactions when we discuss the universality class of the transition\lj{, see Sec.~\ref{sec:discussion}}.

The full action describing both the situations with and without a finite trigonal warping term is hence given by
\begin{align}
  S = \int \rmd\tau\kern.1em\rmd^2\vec{x} \left( \mathcal{L}_0 + \mathcal{L}_{\text{int}} \right).
\end{align}
We conclude this section by reading off the mass dimensions of the quantities appearing in the theory. In the QBT limit, we wish to renormalize the fields such that the coefficient $f_2$ in front of the QBT term remains fixed during the RG.
Then, in the noninteracting limit, the dynamical critical exponent $z = 2$. Consequently, the linear coefficient has mass dimension $[f_1] = 1$ and is RG relevant, while the cubic coefficient is RG irrelevant with $[f_3] = -1$. The four-fermion coupling becomes dimensionless, $[g] = 0$, i.e., the interaction is marginal at tree level.
In the opposite Dirac limit, the renormalization scheme should fix the coefficient $f_1$ of the linear term.
Hence, in this case $z = 1$, $[f_2] = -1$, $[f_3] = -2$, and the four-fermion coupling becomes irrelevant, $[g] = -1$.


\section{Mean-field theory}
\label{sec:mean-field}

We start by discussing the large-$N$ limit, which can be solved exactly in the framework of mean-field theory.
To distinguish the ordered from the disordered phase, it is useful to think in terms of the composite field $\phi \propto \psi^{\dagger}_i(\sigma^3 \otimes \sigma^3) \psi^i$. Then, the symmetric phase corresponds to $\langle \phi \rangle = 0$, while long-range order is characterized by $\phi$ developing a finite vacuum expectation value via spontaneous symmetry breaking. A finite $\langle \phi \rangle \neq 0$ acts as an effective mass term and opens up a full gap in the fermion spectrum. 
For the present interaction channel, the new ground state spontaneously breaks inversion symmetry between the layers.
We rewrite the action solely in terms of $\phi$ by performing a Hubbard-Stratonovich transformation and then carrying out the integral over the fermion fields. This results in an effective action,
\begin{align}
	S_\text{eff}[\phi] & = \int \rmd\tau\kern.1em\rmd^2\vec x \, \tfrac{1}{2}\phi^2 
	\nonumber \\ & \quad
	- \operatorname{Tr}\ln \left[\partial_\tau + \mathcal{H}_0(-\rmi \nabla) - \sqrt{g}\kern.1em\phi \left(\sigma^3 \otimes \sigma^3\right) \right],
\label{eq:actionphi}
\end{align}
where the trace $\operatorname{Tr}(\,\cdot\,)$ is taken over spinor and flavor indices as well as coordinate space.
A meaningful large-$N$ limit is obtained by fixing $g \Nf = \textnormal{const.}$ and $\phi^2/\Nf = \textnormal{const.}$ 
We reiterate that the fermion flavor number $\Nf$ is equivalent to the number of QBT points $N/2$ in the limit of vanishing trigonal warping, while $\Nf = 2N$ when each QBT point splits into four Dirac cones.
From the trace over the flavor indices, the action \eqref{eq:actionphi} for $\phi$ attains an overall factor of $N_\text{f}$. In the large-$N$ limit, the path integral over $\phi$ is then dominated by the extremum of $S_\text{eff}[\phi]$. We assume constant field configurations $\phi(x) \equiv \phi = \text{const.}$, leading to the effective potential $V_\text{eff}(\phi) = \mathcal{V}\, S_\text{eff}[\phi]\bigr|_{\phi(x) = \phi}$, where $\mathcal{V}$ is the spacetime volume.
The mean-field analysis then boils down to minimizing $V_\text{eff}(\phi)$. It proves to be technically more convenient to evaluate $V_{\text{eff}}^\prime(\phi)$ by differentiating \eqref{eq:actionphi} once with respect to $\phi$ and performing the trace over the spinor and flavor indices, yielding in momentum space
\begin{align}
	V_{\text{eff}}^\prime(\phi) = \phi - 4N_{\text{f}}\!\int\frac{\rmd\omega\kern.1em\rmd^2 \vec p}{(2\pi)^3}\frac{g\kern.1em\phi}{\omega^2 + \mathcal{H}_0(\vec{p})^2 + g \phi^2}.
\end{align}
The divergence occurring for large frequency $\omega$ and large momentum $\vec p$ is handled by introducing a finite ultraviolet cutoff $\Lambda$.
In the QBT case, we implement this as the restriction $|\omega| \leqslant f_2 \Lambda^2$ and $|\vec{p}| \leqslant \Lambda$, in agreement with the dynamic scaling exponent $z = 2$ for $f_1 = 0$. By contrast, in the Dirac limit for finite $f_1$, the integral is regularized as $\sqrt{\omega^2 + f_1^2 \vec{p}^2} \leqslant \lvert f_1\rvert \, \Lambda$, respecting the different dynamic exponent $z=1$ for $f_2 = f_3 = 0$.

Let us first recapitulate the case of pure QBT with $\Nf = N/2$ and $f_1 = f_3 = 0$~\cite{ref6}. Then, $\mathcal{H}_0(\vec{p})^2 = f_2^2 \vec{p}^4$ and the integral is soluble in terms of standard functions. Expanding around $\Lambda \to \infty$ and rescaling $\phi/(\sqrt{f_2} \Lambda^2) \mapsto \phi$ and $g/f_2 \mapsto g$, one finds
\begin{align}
	V_{\text{eff}}^\prime(\phi) \propto \phi\left[1 + \frac{gN}{8\pi}\ln\!\left(\tfrac{1}{4}g\phi^2\right)\right].
\end{align}
Thus, the minimum for \emph{all} $g > 0$ is located not at $\phi = 0$, but at the new minimum
\begin{align}
	\phi_0 = 2 g^{-1/2} \rme^{-4\pi/(gN)}.
	\label{eq:vev-mft}
\end{align}
Hence, infinitesimal $g$ leads to spontaneous symmetry breaking for a rotationally invariant QBT, in agreement with the various previous works on the subject \cite{ref6, zhang10, vafekyang, vafekbig, uebelacker11, lang12, scherer12, Cvetkovic, song12}.
We next investigate the stability of the symmetry-broken phase under perturbation by an infinitesimal Dirac term, realized by switching on small non-vanishing $f_1$. To this end, we consider the curvature of the effective potential around $\phi_0$ and expand it in powers of $|f_1/(f_2 \Lambda)| \ll \phi_0$, 
yielding
\begin{align}
	V_{\text{eff}}^{\prime\prime}(\phi_0) \propto \frac{gN}{4\pi}\left[1 - \rme^{8\pi/(gN)}(f_1/4f_2)^{4}\right],
\label{eq:QBTMFTstability}
\end{align}
where we have rescaled $f_1/(f_2\kern.1em \Lambda) \mapsto f_1/f_2$.
The ordered phase is stable (or at least metastable) as long as $V_{\text{eff}}^{\prime\prime}(\phi_0) > 0$. 
For a given fixed and small $f_1$, this condition holds if and only if $g > g_\mathrm{c}$, with the critical coupling
\begin{align}
	g_\mathrm{c}(f_1/f_2) \simeq \frac{2\pi}{N} \left( -\ln\left\lvert \frac{f_1/f_2}{4} \right \rvert \right)^{-1}
\label{eq:QBTMFTphasebound}
\end{align}
valid for $|f_1/f_2| \ll 1$.
The inclusion of $f_3$ is possible as well by numerical means. However, within mean-field theory, this does not lead to qualitatively new physics since the crucial self-energy corrections, which are prerequisite to obtaining a finite $g_\mathrm{c}$ in the QBT limit $f_1/f_2 = 0$, are suppressed for large $N$ (as we shall show explicitly later). At the mean-field level, therefore, spontaneous symmetry breaking occurs for $g > g_\mathrm{c} \geqslant 0$ with $g_\mathrm{c} \to 0$ for $f_1/f_2 \to 0$.

\begin{figure}
\includegraphics[scale=1]{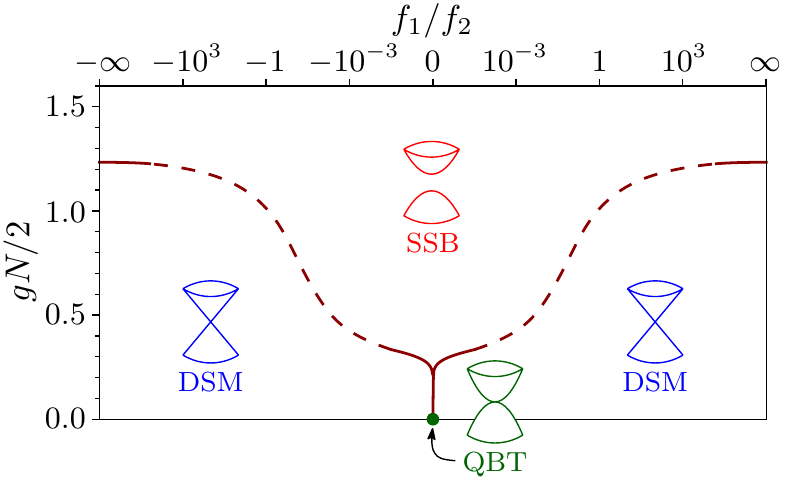}
\caption{
	Mean-field phase diagram for $f_{3} = 0$ using Eqs.~\eqref{eq:QBTMFTphasebound} and \eqref{eq:GNMFTsep}, showing the Dirac semimetal (DSM) phase for small $g < g_\mathrm{c}$ and finite $f_1/f_2$ and the spontaneous-symmetry-broken (SSB) phase for $g > g_\mathrm{c}$. 
	At the origin, $(f_1/f_2, g) = (0,0)$, the fermion spectrum exhibits a quadratic band touching (QBT).
	The dashed curve at intermediate $|f_1 / f_2| \simeq 1$ is given as a guide to the eye.}
\label{fig:MFTphasediag}
\end{figure}

We now turn to the opposite limit of $N_{\text{f}} = 2N$ Dirac flavors perturbed by a small $f_2$ term. We investigate an instability towards the symmetry-broken state by studying the curvature $V_{\text{eff}}^{\prime\prime}(0)$ at the origin $\phi = 0$. 
Since $g$ now has mass dimension $[g] = z - 2 = -1$, we rescale $g \Lambda/f_1 \mapsto g$, $\phi/(\sqrt{f_1}\Lambda^{3/2}) \mapsto \phi$, and $f_1/(f_2 \Lambda) \mapsto f_1/f_2$. To the leading nonvanishing order in $(f_1/f_2)^{-1}$, the curvature is
\begin{align}
	V_{\text{eff}}^{\prime\prime}(0) \propto 1 - \frac{4gN}{\pi^2} \left[1 + \frac{8}{63}\left(\frac{f_1}{f_2}\right)^{\!\!-2}\right].
\end{align}
The phase boundary occurs when the curvature of the effective potential at the origin vanishes, yielding
\begin{align}
	g_{\text{c}}(f_1/f_2) \simeq
	\frac{\pi^2}{4N} \left[1 - \frac{8}{63}\left(\frac{f_1}{f_2}\right)^{\!\!-2} \right],
\label{eq:GNMFTsep}
\end{align}
valid for $|f_1/f_2| \gg 1$.
We note that, within our continuum field theory, the two limiting cases $f_1/f_2 \to \infty$ and $f_1/f_2 \to -\infty$ are in fact equivalent, as they are related by momentum inversion $\vec p \mapsto - \vec p$.
Equation~\eqref{eq:GNMFTsep} in this limit precisely agrees with the known large-$N$ critical coupling in the relativistic Gross-Neveu theory with $2N$ four-component Dirac flavors in $2+1$ dimensions~\cite{hands93, braun11}.
The perturbation~$\propto (f_1/f_2)^{-2}$ is new and represents the influence of the quadratic term in the dispersion $\varepsilon(\vec p)$. It decreases the critical coupling, which is consistent with the general expectation that an increase in the density of states tends to destabilize the disordered semimetallic state.
The combined mean-field phase diagram, showing the phase boundaries both for $|f_1/f_2| \ll 1$ in the QBT regime as well as for $|f_1/f_2| \gg 1$ in the Dirac regime, is depicted in Fig.~\ref{fig:MFTphasediag}.


\section{Renormalization-group analysis}
\label{sec:rg}


\subsection{Flow equations}

\begin{figure}[t]
\centering
\includegraphics[scale=.9]{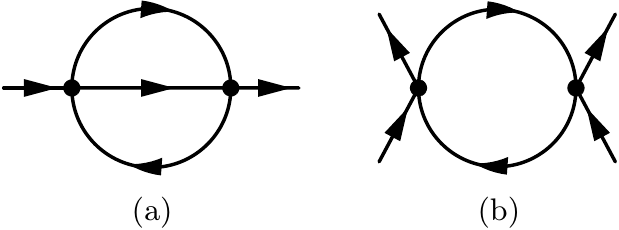}
\caption{Feynman diagrams representing the first non-vanishing loop corrections to self-energy and the four-fermion vertex.}
\label{fig:loops}
\end{figure}

To go beyond the mean-field level, we now turn to an RG analysis. Since two-loop corrections constitute an essential part of the physics we are interested in, we perform field-theoretic renormalization. In our gapless model, the loop integrals will not only have the usual ultraviolet divergences, but also infrared divergences. We regularize these by introducing both an ultraviolet cutoff $\Lambda$ as well as an infrared cutoff $\lambda$, with $\lambda \ll \Lambda$.
For the RG flow, we demand the invariance of the one-particle irreducible effective action $\Gamma$ under the RG step $\lambda \to \lambda/b$ while holding $\Lambda$ fixed, to wit:
\begin{align} \label{eq:effective-action}
	\partial_t \Gamma = 0,
\end{align}
where $\partial_t \equiv \partial/(\partial \ln b)$ with the RG ``time'' $t \equiv \ln b \in [0,\infty)$.
We expand the functional integral within perturbation theory in the coupling $g$. The condition~\eqref{eq:effective-action} allows us to compute $\beta$ functions, which characterize the scale dependence of $g$ and the parameters $f_1$, $f_2$, and $f_3$ within the effective low-energy description.
At tree level, $\Gamma = S$, so that one obtains the canonical scaling dimensions determined at the end of Sec.~\ref{sec:action}. 
For quantum corrections, we work to the leading non-vanishing loop order. The self-energy and four-point vertex correction diagrams are shown in Fig.~\ref{fig:loops}. 
Possible diagrams of the same order that are not shown vanish in the the present single-channel approximation. Note that the one-loop self-energy diagram is absent due to kinematics: It is independent of the external momentum and hence can at most generate a mass term, which is forbidden by symmetry.

We briefly sketch the general strategy regarding the evaluation of the loop corrections, beginning with the (evidently more challenging) two-loop self-energy correction shown in Fig.~\ref{fig:loops}(a). The diagram has the so-called {\it sunset} topology. For translationally invariant systems, the evaluation of such diagrams turns out to be particularly efficient when carried out in position space~\cite{sunrisexspace}. This way, the evaluation of the diagram ultimately leads to a single position-space integral, which is a considerable technical simplification over the corresponding momentum-space version. In particular, the two-loop contribution to the self-energy simply becomes
\begin{align}
	\Sigma(\omega,\vec p) &\propto
	\int \rmd\tau\,\rmd^2\vec{x}\, \rme^{-\rmi (\omega \tau + \vec p \cdot \vec x)}
	\nonumber\\
	&\quad{}\times \left(\sigma^3\otimes\sigma^3\right) G_0(\tau,\vec x)\left(\sigma^3\otimes\sigma^3\right)
	\\
	&\quad{}\times G_0(-\tau,-\vec x)\left(\sigma^3\otimes\sigma^3\right)G_0(\tau,\vec x)\left(\sigma^3\otimes\sigma^3\right), \nonumber
\end{align}
where $G_0(\tau,\vec x) = [\partial_\tau + \mathcal H_0(-\rmi \nabla)]^{-1}$ is the tree-level propagator and we have suppressed coupling constants, numerical prefactors, and contraction over spinor indices for brevity.
Similarly, the contribution to the four-fermion vertex can be written in position space as
\begin{multline}
	\delta \Gamma^{(4)} \propto \int\rmd\tau\,\rmd^2\vec{x}\, G_0(\tau,\vec x)\left(\sigma^3\otimes\sigma^3\right)\\ \times G_0(-\tau,-\vec x)\left(\sigma^3\otimes\sigma^3\right),
\end{multline}
where we have also set external coordinates to zero.
All position-space integrals are regularized by short-wavelength $\ell \propto 1/\Lambda \simeq a$ and long-wavelength $L \propto 1/\lambda$ cutoffs.
In the QBT regime, we restrict the domain of integration as $\pi/(2\Lambda) \leqslant |\vec x| \leqslant \pi/(2\lambda)$ with unrestricted integration over $\tau$. In the Dirac regime, we choose $\pi/(2\Lambda) \leqslant \sqrt{\tau^2 + {\vec x}^2} \leqslant \pi/(2\lambda)$.
This regularization prescription not only respects the different symmetries in the strict QBT and Dirac limits, but also allows us to perform the loop integrations analytically. The {\it a priori} arbitrary constant $\pi/2$ in the definition of the position-space cutoffs $\ell = \pi/(2\Lambda)$ and $L = \pi/(2\lambda)$ has been chosen such that the resulting large-$N$ critical coupling $g_\mathrm{c}$ in the Dirac limit $|f_1/f_2| \to \infty$ matches the mean-field result, Eq.~\eqref{eq:GNMFTsep}. %
Evidently, a major part of the difficulty of evaluating the loop corrections now resides in computing the position space propagator $G_0(\tau,\vec{x})$, additional information pertaining to which is given in Appendix~\ref{app:loops}.

The fact that the location and shape of the Fermi surface changes when a QBT point splits into four Dirac cones requires us to start with separate discussions of the two cases $|f_1/f_2| \ll 1$ and $|f_1/f_2| \gg 1$.
To obtain the full RG flow also for finite values of $|f_1 / f_2| \sim 1$, we shall eventually interpolate between the respective limits by means of a suitable Pad\'e approximation.

Let us start by discussing the QBT limit, to which an infinitesimal $|f_1/f_2| \ll 1$ has been added. Keeping the quadratic coefficient $f_2$ fixed, the $\beta$ function $\beta(g) \equiv \partial_t g \equiv \partial g / (\partial \ln b)$ for the dimensionless short-range interaction $g$ becomes
\begin{multline}
	\beta(g) = \frac{g^2(\NQ-1)}{2\pi}\left[ 1 - \frac{7 - 4\ln 2}{240}\frac{\pi^2}{4}\left(\frac{f_1}{f_2}\right)^2 
	\right. \\ \left.
	{}+ \frac{1}{2}\left(\frac{f_1}{f_2}\right)\!\left(\frac{f_{3}}{f_2}\right) + \frac{32}{\pi^2}\left(\frac{f_3}{f_2}\right)^2 \right],
\label{eq:betagQBT}
\end{multline}
where we have rescaled $f_1/(f_2\Lambda) \mapsto f_1/f_2$, $f_3 \Lambda/f_2 \mapsto f_3/f_2$, and $g/f_2 \mapsto g$ as in the mean-field theory. We note that the above equation is valid only for $N > 1$. For $N=1$, there is an additional Fierz identity which leads to a finite $\beta$ function for $g$ in this case as well.
The self-energy diagram in Fig.~\ref{fig:loops}(a) leads to a nontrivial flow of the small parameters $f_1$ and $f_3$. We find, to the leading order in $f_1 / f_2$ and $f_3 / f_2$
\begin{align}
\begin{split}
	\beta\!\left(\frac{f_1}{f_2}\right) &= \left(1 - \frac{g^2(2\NQ-1)}{144\pi^2}\right) \left(\frac{f_1}{f_2}\right) \\
	&\qquad + \frac{11g^2(2\NQ-1)}{54\pi^4} \left(\frac{f_3}{f_2} \right) \label{eq:betafQBT}
\end{split}
\end{align}
and
\begin{multline}
	\beta\!\left(\frac{f_{3}}{f_2}\right) = -\left(1 + \frac{59g^2(2\NQ-1)}{3456\pi^2}\right)\left(\frac{f_3}{f_2}\right)
	\\
	+ \frac{g^2(2\NQ-1)}{576}\left(\frac{f_1}{f_2}\right). \label{eq:betafm1QBT}
\end{multline}
The anomalous field dimension $\eta_\psi$ reads in this limit
\begin{align}
\begin{split}
	\eta_\psi &= \frac{g^2 (2\NQ - 1)}{4\pi^2} \left[\frac{1}{18}
- \frac{25-36\ln\frac43}{2880}\frac{\pi^2}{4}\!\left(\frac{f_1}{f_2}\right)^{\!\!2} \right. \\
& \quad {} \left. -\left(\frac32 \ln\frac{4}{3}-\frac{179}{432}\right)\!\left(\frac{f_1}{f_2}\right)\!\left(\frac{f_3}{f_2}\right) 
+ \frac{871}{243\pi^2}\left(\frac{f_3}{f_2}\right)^{\!\!2}\right] \label{eq:etapsiQBT}
\end{split}
\end{align}
and the dynamic critical exponent $z$ becomes
\begin{multline}
	z = 2 - \frac{g^2 (2\NQ - 1)}{4\pi^2}\left[\frac{9\ln\frac43 - 2}{72} 
	\right. \\ {}
	+ \frac{49-30\ln 2 - 9\ln 3}{5760}\frac{\pi^2}{4}\left(\frac{f_1}{f_2}\right)^{\!\!2} \\ \left.{}- \frac{1593\ln\frac43 - 422}{216}\!\left(\frac{f_1}{f_2}\right)\!\left(\frac{f_3}{f_2}\right)\right.\\
	\left.{}+ \frac{1850 - 4860\ln\frac{4}{3}}{243\pi^2} \left(\frac{f_3}{f_2}\right)^{\!\!2}\right].
 \label{eq:zQBT}
\end{multline}
Note that the contribution $\propto (f_1/f_2)^2$ to $z$ is negative, tending to decrease the dynamic exponent from $z=2$ towards the Dirac value of $z=1$ for large enough $|f_1/f_2|$.

In the Dirac limit with a quadratic perturbation $\propto f_2$ added to the Hamiltonian, the effects of the strongly-irrelevant cubic coefficient $f_3$ can be safely neglected as noted above. We find, for $|f_1/f_2| \gg 1$, $f_3 = 0$, and $\Nf = 2N$ Dirac flavors, for the flow of the short-range interaction
\begin{align}
\beta(g) &= -g + \frac{g^2(4N-1)}{\pi^2}\left[1 + \frac{4128}{35\pi^2}\left(\frac{f_2}{f_1}\right)^{\!\!2} \right], \label{eq:betagDir}
\end{align}
where we now have rescaled $g \Lambda/f_1 \mapsto g$ and $f_2 \Lambda / f_1 \mapsto f_2/f_1$.
The small parameter $f_2/f_1$ is irrelevant in the Dirac limit. Its flow reads
\begin{align}
\beta\!\left(\frac{f_2}{f_1}\right) &= -\left[1 + \frac{29}{420}\frac{g^2(8N-1)}{\pi^4} \right]\!\left(\frac{f_2}{f_1}\right). \label{eq:betafDir}
\end{align}
The anomalous dimension $\eta_\psi$ and the dynamic critical exponent $z$ become in this limit
\begin{align}
\eta_\psi &= \frac{g^2(8N-1)}{\pi^4}\left[\frac{1}{12} + \frac{1312}{105\pi^2}\left(\frac{f_2}{f_1}\right)^{\!\!2} \right] \label{eq:etapsiDir},\\
z &= 1 - \frac{8g^2(8N-1)}{\pi^6}\left(\frac{f_2}{f_1}\right)^{\!\!2}. \label{eq:zDir}
\end{align}
%

\subsection{Basic flow properties}
\label{subsec:discuss}

Before solving the full set of flow equations to construct phase diagrams, we first extract some general characteristics by analytical means. We begin with the Dirac case, which in the limit $|f_1/f_2| \to \infty$ boils down to the $(2+1)$-dimensional relativistic Gross-Neveu theory.
Apart from the fully attractive noninteracting Dirac fixed point
\begin{align}
\text{D}: \qquad \left(f_1/f_2, g\right)_* = \left(\pm \infty,0 \right),
\end{align}
the only interacting fixed point for $|f_1/f_2| \gg 1$ is at
\begin{align}
\text{GN}_3: \qquad \left(f_1/f_2, g \right)_* = \left(\pm \infty, \frac{\pi^2}{4N-1}\right).
\end{align}
The fixed point $\text{GN}_3$ is characterized by a dynamic critical exponent $z=1$ and an anomalous dimension 
\begin{align}
	\eta_\psi = \frac{8N-1}{12(4N-1)^2}.
\end{align}
For $N = 2$, this yields $\eta_\psi = 0.026$.
Within our approximation, the correlation-length exponent $\nu = 1$, but there will be $N$-dependent corrections once higher loop orders are taken into account.
GN$_3$ has a unique RG relevant direction along the $g$ axis, 
\lj{as $f_2/f_1$ is irrelevant in its vicinity. We also note that other short-range interactions, such as flavor-symmetry-breaking operators, are irrelevant at this fixed point~\cite{gehring15}. GN$_3$} describes a transition from the semimetallic Dirac phase for $g<g_*$ to an ordered phase for $g>g_*$, in which the fermions acquire a dynamical mass gap as a consequence of spontaneous symmetry breaking. Hence, GN$_3$ is an incarnation of the celebrated relativistic Gross-Neveu critical point in $2+1$ dimensions~\cite{hands93, vasiliev93, gracey94, vojta00a, vojta00b, braun11, gracey16, mihaila17, zerf17, iliesiu18, ihrig18}.
In our interaction channel, the ordered state is characterized by $\langle \psi^\dagger (\sigma^3 \otimes \sigma^3) \psi\rangle \neq 0$, which spontaneously breaks inversion symmetry between the layers~\cite{vafekbig}.
In the large-$N$ limit, the Gross-Neveu fixed-point value is $g_* = \pi^2/(4N) + \mathcal{O}(1/N^2)$, in agreement with the result we found in the mean-field theory, Eq.~\eqref{eq:GNMFTsep}. 
Note that values of couplings are in principle nonuniversal and depend on the regularization scheme. Here, we have adapted our position-space regularization to match the mean-field result for the critical coupling. However, we emphasize that this agreement may not carry over in the case of other nonuniversal quantities.
For instance, this is the case for the separatrix that defines the phase boundary between the Dirac semimetal and the interaction-induced insulator for $|f_1/f_2| \gg 1$, which is obtained from the RG flow as
\begin{align}
	g_{\text{c}}(f_1/f_2)
	\simeq \frac{\pi^2}{4N-1} \left[1 - \frac{4128}{35\pi^2}\left(\frac{f_1}{f_2} \right)^{\!\!-2} \right].
\end{align}
which is in qualitative, but not quantitative, agreement with the mean-field result, Eq.~\eqref{eq:GNMFTsep}. 
We reiterate that this discrepancy is a consequence of the difference in regularization schemes and has no effect on universal observable quantities such as critical exponents, mass ratios, etc., which are regularization independent.

We now proceed to the QBT limit for $0 \leqslant |f_1/f_2| \ll 1$. In this regime, there is only the Gaussian fixed point at 
\begin{align}
\text{Q}: \qquad (f_1/f_2,g)_* = \left(0, 0\right),
\end{align}
describing a noninteracting Fermi system with a quadratic dispersion.
Q has a marginal direction along the $g$ axis, while $f_1$ is power-counting relevant.

Let us first review the $\mathrm{O}(2)$-invariant case for $f_1 = f_3 = 0$ and $g>0$ in order to connect with the previous work \cite{ref6}.
In this case, $\beta(g)$ is positive and finite for all $g>0$, implying an instability of the system towards the infrared. More precisely, integrating the RG flow equation for $g$, Eq.~\eqref{eq:betagQBT}, we find for $f_1 = f_3 = 0$,
\begin{align}
  g(t) = \frac{1}{1/g_0 - t (N-1)/(2\pi)}, \qquad \text{for } t \leqslant t_\text{SSB},
  \label{eq:timeevg}
\end{align}
with initial value $g_0 \equiv g(t=0)$. Patently, the evolution of $g$ exhibits a pole at a finite RG time $t_\text{SSB} = 2\pi/[g_0(\NQ - 1)]$.
Informed by the mean-field analysis, we can trace back this runaway flow to an instability of the semimetallic state towards the interaction-induced insulator. The latter is characterized by inversion-symmetry breaking and a finite vacuum expectation value of the fermion bilinear $\langle \phi \rangle \propto \langle \psi^\dagger_i (\sigma^3 \otimes \sigma^3) \psi^i \rangle \neq 0$.
Identifying the corresponding energy scale $\lambda_\text{SSB}^2 \simeq \Lambda^2\,\rme^{-2t_\text{SSB}}$, at which the instability occurs, allows us to estimate the effective amplitude of the condensate, yielding
\begin{align}
\langle \phi \rangle \propto \lambda_\text{SSB}^2 \propto \rme^{-4\pi/[g_0(\NQ - 1)]},
\end{align}
where we have used the order parameter's scaling dimension $[\phi] = (z+2)/2 = 2$.
It is conceptually satisfying to note that the exponential factor in the above estimate in the limit $N \to \infty$ agrees precisely with the mean-field result, Eq.~\eqref{eq:vev-mft}. This furnishes a nontrivial consistency check.

The RG flow equations also permit to compute the form of the phase boundary at finite $0 < |f_1/f_2| \ll 1$. To this end, we consider trajectories in parameter space starting infinitesimally close to the non-interacting QBT fixed point Q. In this regime, $f_1/f_2$ flows according to its canonical scaling dimension, $(f_1/f_2)(t) = (f_1/f_2)_0\,\rme^{-t}$, where $(f_1/f_2)_0 \equiv (f_1/f_2)(t=0)$, whereas the RG evolution of $g$ is given by Eq.~\eqref{eq:timeevg} above. Eliminating the RG time $t$, one finds the RG trajectories near the fixed point Q as
\begin{align}
	g(f_1/f_2) = \frac{2\pi}{N-1} \frac{1}{ \ln C - \ln|f_1/f_2| }
\label{eq:trajectory}
\end{align}
with a positive constant $C = \rme^{2\pi/\left[g_0(\NQ-1)\right]} |f_1/f_2|_0$ that is determined by the initial values $\left((f_1/f_2)_0, g_0 \right)$ of the flow for $t=0$.
Each member of the family of RG trajectories defined by Eq.~\eqref{eq:trajectory} and parametrized by $C$ can now be continued ``backwards'' in RG flow for $t \to - \infty$ and will eventually approach the noninteracting QBT fixed point Q.
In the opposite RG time direction, $t \to \infty$, one member of the family must be the separatrix that precisely flows into the critical Gross-Neveu fixed point GN$_3$ in the Dirac limit for $|f_1/f_2| \gg 1$.
In the mean-field theory, this happens for $C = \ln 4$, for which the large-$N$ limit of Eq.~\eqref{eq:trajectory} agrees with Eq.~\eqref{eq:QBTMFTphasebound}.
Without the mean-field input, the perturbative RG analysis around Q for $|f_1/f_2| \ll 1$ alone has nothing to say on which of the trajectories is the separatrix; we shall discuss in the following subsection how to circumvent this problem by making use of the flow near the Gross-Neveu fixed point in the opposite limit $|f_1/f_2| \gg 1$.
In this subsection, we suffice ourselves by noting that a separatrix that connects Q with GN$_3$ exists for all $N$ and has the form as given in Eq.~\eqref{eq:trajectory}.

Let us now discuss the situation for $f_3 \neq 0$, which induces nontrivial self-energy corrections that go beyond the mean-field result.
To see this, consider the trajectories starting on the $f_1 = 0$ line. Inspecting the flow equations, one finds the slope of all trajectories with $g > 0$ as
\begin{align}
	\frac{\rmd (f_1/f_2)}{\rmd g} 
	& = \frac{\beta(f_{1}/f_2)}{\beta(g)} 
	\nonumber \\
	& = \frac{11}{27\pi^3}\frac{2\NQ-1}{\NQ-1}\frac{f_3}{f_2}\frac{1}{1+\frac{32}{\pi^2}(f_3/f_2)^2} + \mathcal O(f_1/f_2).
\end{align}
Importantly, the slope is finite for all $g>0$, implying that every RG trajectory (except the one that connects the free theories Q and D at fixed $g=0$) crosses the line $f_1 = 0$ at a finite $g$ when $f_3 \neq 0$. In particular, this is true for the separatrix that connects Q with GN$_3$. There must, therefore, be a critical coupling strength $g_\text{c} > 0$, below which the system flows to a Dirac semimetal phase. This hence provides a rigorous RG demonstration of the phenomenology found numerically in Ref.~\cite{Pujarimain}.
It is also consistent with the result obtained recently within a random phase approximation~\cite{honerkamp17b}.

We close by answering why the mean-field theory is unaware of this behavior, the reason for which is more transparent when the above is expressed in terms of the 't Hooft coupling $g' \equiv g\NQ$, which remains finite at the Gross-Neveu fixed point GN$_3$. In $(f_1/f_2,g')$ space, the same slope is
\begin{align}
	\frac{\rmd (f_1/f_2)}{\rmd g'} = \frac{22}{27\pi^3 \NQ} \frac{f_3}{f_2} + \mathcal O\!\left(f_1/f_2,1/\NQ^2\right),
	\label{eq:slope-large-N}
\end{align}
and is therefore subleading when sending $\NQ \to \infty$ while keeping $g' = \text{const.}$
In other words, self-energy effects are suppressed in the large-$N$ limit.


\subsection{Phase diagrams}
\label{subsec:phaseportrait}

\begin{figure*}
\includegraphics[width=\textwidth]{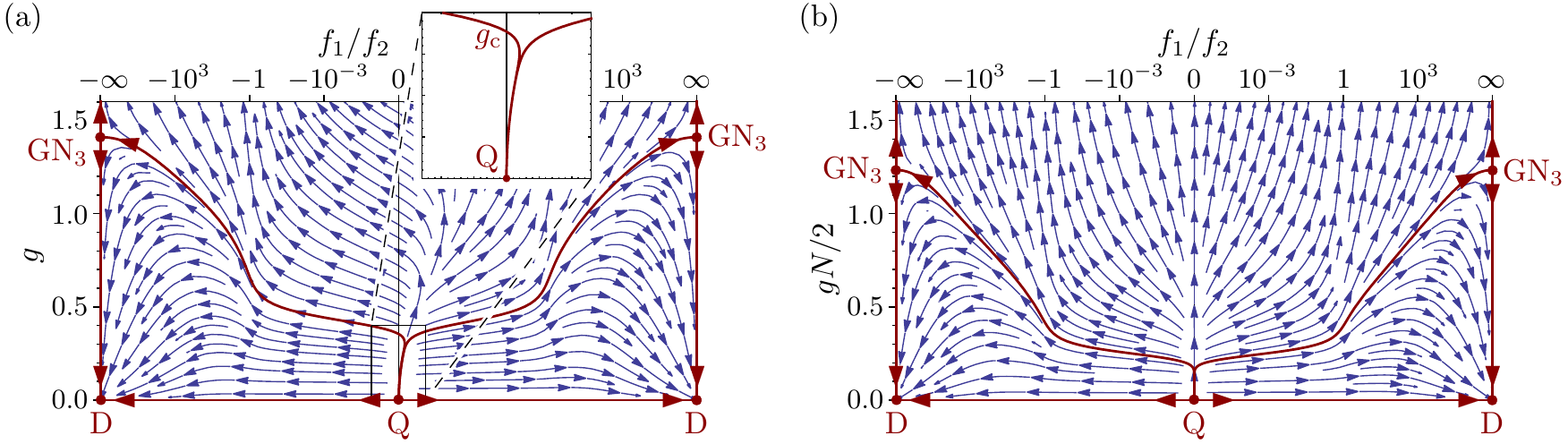}
\caption{
	RG flow diagram in the plane spanned by $(f_1/f_2)$ and $g$ for constant $f_3 = (2\sqrt{3})^{-1}$ for (a) $N=2$ and (b) in the limit $N \to \infty$.
	The Gaussian fixed points corresponding to the non-interacting Dirac and QBT systems are denoted D and Q respectively, while GN$_3$ is the $(2+1)$-dimensional Gross-Neveu fixed point.
	The separatrices connecting the different fixed points are shown in dark red. 
}
\label{fig:RGphaseport1}
\end{figure*}

We proceed to construct the RG phase diagram in the full $(f_1/f_2, g)$ coupling space. 
As the configuration of the Fermi surface changes from the QBT limit for $|f_1/f_2| \ll 1$ to the Dirac limit for $|f_1/f_2| \gg 1$, the standard regularization scheme in momentum space, as well as our position-space regularization scheme, required us to {\it a priori} treat these different regimes separately.
This approach led us to the flow equations \eqref{eq:betagQBT}--\eqref{eq:betafm1QBT} in the former limit and \eqref{eq:betagDir}--\eqref{eq:betafDir} in the latter, and these equations should be understood as asymptotic expansions in the two different regimes of an unknown set of flow equations valid for all $f_1/f_2$.
A useful approximation to these can be obtained by employing suitable Pad\'e approximants which interpolate between the known limits.
The $[m/n]$ Pad\'e approximant is defined as a degree $m$/degree $n$ rational function, where the coefficients in the polynomial numerator and denominator are chosen such that the approximant reproduces the correct expansions for $|f_1/f_2| \ll 1$ (QBT regime) and $|f_1/f_2| \gg 1$ (Dirac regime).
For the flow equations of $f_1/f_2$ and $f_3/f_2$ we use $[3/2]$ and $[2/2]$ Pad\'e approximants, 
\begin{align}
	\beta (f_1/f_2) &= \frac{a_0 + a_1 (f_1/f_2) + a_2 (f_1/f_2)^2 + (f_1/f_2)^3}{1 + b_1 (f_1/f_2) + b_2 (f_1/f_2)^2}, \label{eq:pade-1} \\
	\beta (f_3/f_2) &= \frac{c_0 + c_1 (f_1/f_2) + c_2 (f_1/f_2)^2}{1 + (f_1/f_2)^2}, \label{eq:pade-2}
\end{align}
which corresponds to the minimal degree necessary to match Eqs.~\eqref{eq:betafQBT}, \eqref{eq:betafm1QBT}, and \eqref{eq:betafDir}.
Other choices are in principle possible as well, and the above approximants have been selected under the demand that they be of minimal degree needed to faithfully reproduce the asymptotic expansions for $f_1/f_2 \to 0$ and $f_2/f_1 \to 0$, respectively, and do not introduce any unphysical poles in the resulting Pad\'e-approximated flow equations.
For the flow equation of $g$, it proved advantageous to perform the interpolation separately for the even and odd parts in $f_1/f_2$, explicitly
\begin{align}  \label{eq:pade-3}
	\beta(g) = \frac{d_0 + d_2 (f_1/f_2)^2 + d_4 (f_1/f_2)^4}{1 + e_2 (f_1/f_2)^2 + (f_1/f_2)^4} + \frac{d_1 (f_1/f_2)}{1 + (f_1/f_2)^4}.
\end{align}
Note that the coefficients $a_i$, $b_i$, $c_i$, $d_i$, and $e_i$ are independent of $f_1/f_2$, but depend on $g$ and $f_3/f_2$.
Their explicit values are given in Appendix~\ref{app:pade}.

The resulting RG flow diagram for $\NQ = 2$, relevant for the honeycomb bilayer, is depicted in Fig.~\ref{fig:RGphaseport1}(a). 
The diagram shows a cut through parameter space at a fixed $f_3/f_2 = (2\sqrt{3})^{-1}$ in the QBT regime, chosen to match the microscopic tight-binding value in the honeycomb bilayer, Eq.~\eqref{eq:fm10fromTB}.
For simplicity, we have identified here the ultraviolet cutoff $\Lambda$ with the inverse of the lattice constant.
Besides the noninteracting fixed points Q at $f_1/f_2 = 0$ and D at $|f_1/f_2| = \infty$, the critical Gross-Neveu fixed point GN$_3$ at $|f_1/f_2| = \infty$ is the only interacting fixed point. We reiterate that the two vertical axes at $f_1/f_2 = + \infty$ and $f_1/f_2 = - \infty$ should be identified with each other, as they are related by inversion symmetry $\vec p \mapsto - \vec p$ emerging for $f_2 = f_3 = 0$.
All RG trajectories for $g>0$ cross the QBT axis at $f_1/f_2 = 0$ with a finite slope. The separatrix connecting Q with GN$_3$ in the regime $f_1/f_2 \geqslant 0$ therefore crosses this line at a finite value of the coupling.
The critical coupling $g_\text{c}$ at which this happens for $N=2$ and fixed $f_3/f_2 = (2\sqrt{3})^{-1}$ in the QBT regime is found to be $g_\text{c} \approx 0.35$.
We have checked numerically that the inclusion of the running of $f_3/f_2$ in the QBT regime does not change the qualitative characteristics of the flow diagram, and only moderately modifies its quantitative features. In particular, we find that the improved critical coupling that includes the running of $f_3/f_2$ is $g_\text{c} \approx 0.40$ for the initial value $(f_3/f_2)(t=0) = (2\sqrt{3})^{-1}$.

This should be contrasted with the situation for $N \to \infty$, depicted in Fig.~\ref{fig:RGphaseport1}(b). In this limit, the flow diagram becomes symmetric with respect to $f_1/f_2 \mapsto - f_1/f_2$, and the separatrices no longer cross the QBT axis for strict $N = \infty$. Inclusion of a finite $f_3/f_2$ has qualitatively no influence. In the QBT limit, the critical coupling $g_\text{c}$, below which the semimetallic phase is stable, vanishes for large $N$, implying spontaneous symmetry breaking for all finite values of $g > 0$.
As an aside, we note the qualitative agreement between Figs.~\ref{fig:RGphaseport1}(b) and \ref{fig:MFTphasediag}, which is reassuring.

\begin{figure}[t]
\includegraphics[width=\linewidth]{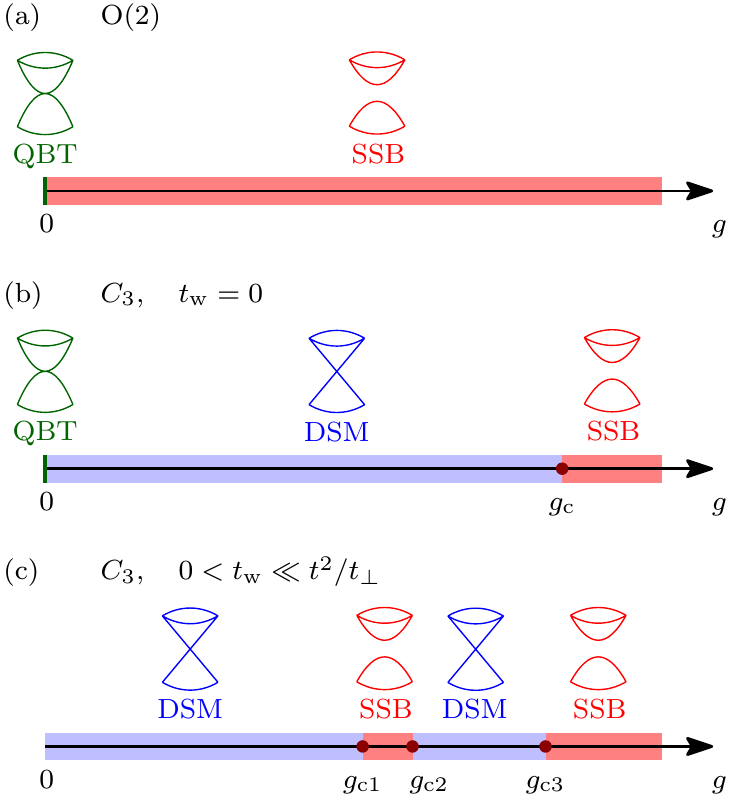}
\caption{Schematic low-temperature phase diagram of QBT systems with 
	(a) full rotational $\mathrm{O}(2)$ symmetry, 
	(b) $C_3$ symmetry without trigonal warping $t_\text{w} = 0$ and 
	(c) sufficiently small trigonal warping $0 < t_\text{w}  \ll t^2/t_\perp$ on the microscopic level, 
	as a function of the short-range interaction $g$. The insets indicate the low-energy fermion spectra in the quadratic band touching (QBT), Dirac semimetal (DSM), and spontaneous-symmetry-broken (SSB) phases.
}
\label{fig:phase-diag-f1-f2-f3}
\end{figure}

The low-temperature physics conveyed by the RG flow can be summarized as follows:
\begin{enumerate}[(a)]
\item For initial couplings $f_1 = f_3 = 0$, which corresponds to the QBT with the full rotational $\mathrm{O}(2)$ symmetry, there is an instability already at infinitesimal coupling, in agreement with the previous works~\cite{ref6, zhang10, vafekyang, vafekbig, uebelacker11, lang12, scherer12, Cvetkovic, song12}, see Fig.~\ref{fig:phase-diag-f1-f2-f3}(a).
\item For the QBT systems with $C_3$ symmetry only and trigonal warping tuned to zero, $f_1 = 0$ and $f_3 \neq 0$, there is a stable semimetallic phase for $g < g_\mathrm{c}$ with a finite critical coupling $g_\mathrm{c} > 0$, see Fig.~\ref{fig:phase-diag-f1-f2-f3}(b). The instability occurs only for $g>g_\mathrm{c}$, in agreement with the numerics of Ref.~\cite{Pujarimain}. The critical coupling vanishes in the large-$N$ limit, as well as when all $\mathrm{O}(2)$-breaking microscopic perturbations, such as $f_3$, vanish.
\item When the QBT point is split into the four symmetry-allowed Dirac points by a sufficiently small positive trigonal warping, a more complex scenario emerges. For initial (microscopic) parameters $0 < f_1/f_2 \ll 1$, lines of constant $f_1/f_2$ cross a separatrix connecting Q and GN$_3$ three times. This leads to a rich phase diagram as a function of the short-range coupling $g$, including three quantum phase transitions at $g_{\mathrm{c}i}$, $i=1,2,3$, between semimetallic and symmetry-broken phases, see Fig.~\ref{fig:phase-diag-f1-f2-f3}(c). In the limit $f_1/f_2 \searrow 0$, both $g_{\mathrm{c}1}$ and $g_{\mathrm{c}2}$ go to zero, and we recover the standard phase diagram comprising a single critical coupling $g_\mathrm{c} \equiv g_{\mathrm{c}3}$.
This scenario is directly testable in current numerical setups~\cite{lang12, Pujarimain}.
\end{enumerate}

\section{Discussion}
\label{sec:discussion}

Let us discuss the critical behavior that should be expected for the continuous semimetal-insulator transitions that we have established in the QBT systems with $C_3$ symmetry.
We first note that all RG fixed points that we have found by interpolating between the QBT and Dirac regimes are located in the strict limits $f_1/f_2 = 0$ and $|f_1/f_2| \to \infty$, respectively. That this must be so, at least on the level of perturbation theory, can be inferred from the following indirect argument. Assume that a fixed point at finite $0 < |f_1/f_2| < \infty$ exists. Such a fixed point would describe a scale-invariant Dirac system in which the coefficient $f_2/f_1$ of the quadratic term in the dispersion does not flow. This, however, is in contradiction with the fact that $f_2/f_1$ is power-counting irrelevant.
Thence, the only possible path in parameter space for the separatrix emanating from the fixed point GN$_3$ when continuing backwards in RG time is through the QBT axis $f_1 / f_2 = 0$ (crossing this axis, as we have seen above, at a finite angle), eventually approaching the fixed point Q.
This general argument is in agreement with our explicit findings, see Fig.~\ref{fig:RGphaseport1}.

The quantum critical transitions shown in Figs.~\ref{fig:phase-diag-f1-f2-f3}(b) and (c) are therefore described by the fully relativistic Gross-Neveu universality class with dynamic exponent $z=1$, comparatively large correlation-length exponent $\nu = 1 + \mathcal O(1/N)$, and large order-parameter anomalous dimension $\eta_\phi = 1  + \mathcal O(1/N)$.
The $\mathcal O(1/N)$ corrections to these exponents depend on the symmetry of the order parameter and the number of fermion flavors, as we discuss in the following.

For the case of spinless fermions on the honeycomb bilayer, natural instabilities are towards an inversion-symmetry-broken state~\cite{vafekbig}, a charge-density wave \cite{scherer12}, or a quantum anomalous Hall phase \cite{ref6, zhu16, zeng18, sur18}, all of which spontaneously break Ising $\mathbb Z_2$ symmetries. The critical exponents of the corresponding Gross-Neveu-Ising universality class are well-established~\cite{hands93, vasiliev93, gracey94, vojta00a, vojta00b, braun11, gracey16, mihaila17, zerf17, iliesiu18, ihrig18}. Within the $1/N$ expansion, they read~\cite{vasiliev93, gracey94, janssen14}
\begin{align}
	1/\nu & = 1 - \frac{4}{3\pi^2 N} + \frac{632+27\pi^2}{27\pi^4 N^2} + \mathcal O(1/N^3) \nonumber \\
	& \approx 1.018(85) \qquad \text{for } N=2,	\label{eq:nu-GNI}
	\\
	\eta_\phi & = 1 - \frac{8}{3\pi^2 N} + \frac{304-27\pi^2}{27\pi^4 N^2} + \mathcal O(1/N^3) \nonumber \\
	& \approx 0.868(4) \qquad \text{for } N=2.
\end{align}
Here, we have estimated the numerical uncertainty for the $N=2$ case from the size of the $\mathcal O(1/N^2)$ correction.
Note that $N$ in our notation corresponds to the number of QBT points in the microscopic theory, each of which splits into four Dirac points with two-component (pseudo-)spinors (equivalent to two four-component Dirac flavors per QBT valley), in the case without a physical spin.
The values of the other exponents $\alpha$, $\beta$, $\gamma$, and $\delta$ can be obtained from $\nu$ and $\eta_\phi$ by means of the usual hyperscaling relations~\cite{herbutbook}.
For the fermion anomalous dimension, even the $\mathcal O(1/N^3)$ correction is known,
\begin{align}
	\eta_\psi & = \frac{1}{3\pi^2N} + \frac{28}{27\pi^4 N^2} 
	\nonumber \\ & \quad
	- \frac{501+2268 \zeta(3) - \pi^2(94+216 \ln 2)}{1296 \pi^6 N^3} + \mathcal O(1/N^4) \nonumber \\
	& \approx 0.0195(1) \qquad \text{for } N=2.
\end{align}
Although the precise determination of the exponents has not been the focus of our work, it is satisfying to note broad agreement of the above results with our RG calculations, which led to the estimates $\nu \approx 1$ and $\eta_\psi \approx 0.026$ for $N = 2$, as noted earlier.

For the spin-$1/2$ case, the number of fermion flavors is doubled, i.e., $N=4$ for the case of the honeycomb bilayer. An instability towards a charge density wave phase is possible in this case as well upon tuning the nearest-neighbor repulsion~\cite{scherer12}. Such a transition would be described by the Gross-Neveu-Ising universality class with the above equations evaluated for $N=4$, leading to $1/\nu \approx 0.988(21)$, $\eta_\phi \approx 0.933(1)$, and $\eta_\psi = 0.00910(1)$.
The most natural instability, however, which occurs upon tuning the on-site Hubbard repulsion, is towards the N\'eel antiferromagnet~\cite{lang12, Pujarimain}, spontaneously breaking the Heisenberg $\mathrm{SU}(2)$ spin symmetry. The critical behavior of the continuous transition is described by the corresponding Gross-Neveu-Heisenberg universality class~\cite{janssen14, toldin15, otsuka16, zerf17, knorr18, gracey18}. In the $1/N$ expansion, the exponents are~\cite{gracey18}
\begin{align}
	1/\nu & = 1 - \frac{4}{\pi^2 N} + \frac{104+9\pi^2}{3\pi^4 N^2} + \mathcal O(1/N^3) \nonumber \\
	& \approx 0.940(41) \qquad \text{for } N=4,
	\\
	\eta_\phi & = 1 + \frac{16+3\pi^2}{3\pi^4 N^2} + \mathcal O(1/N^3) \nonumber
	\\ &
	\approx 1.010(10) \qquad \text{for } N=4,
\end{align}
and
\begin{align}
	\eta_\psi & = \frac{1}{\pi^2N} + \frac{4}{3\pi^4 N^2} 
	\nonumber \\ & \quad
	- \frac{332-378 \zeta(3) + 9\pi^2(5+4\ln 2)}{72 \pi^6 N^3} + \mathcal O(1/N^4) \nonumber \displaybreak[1] \\
	& \approx 0.0261(1) \qquad \text{for } N=4,
\end{align}
with $z=1$.
We note that the rough estimates  $\nu = 1.0(2)$ and $z=0.9(2)$ obtained in the simulations of spin-$1/2$ fermions on the honeycomb bilayer~\cite{Pujarimain} are consistent with the above values for $N=4$ (corresponding to eight flavors of four-component Dirac spinors).

Let us append a discussion on the expected finite-temperature phase diagram, assuming a QBT system without trigonal warping, $t_\text{w} = 0$, on the microscopic level.
The qualitative finite-temperature behavior can be obtained from the RG by noting that temperature sets a scale at which the flow is effectively cut off.
For weak interactions $g \ll g_\mathrm{c}$, the RG scale at which the flow escapes the regime of fixed point Q is exponentially suppressed, leading to a large regime of temperature values at which the dynamic critical exponent is effectively $z = 2$, see Fig.~\ref{fig:finite-T}. Signatures of the splitting into Dirac cones will only be observable at low temperatures $T  \lesssim (T_*/N^2) \exp({-\frac{4\pi}{g N}})$, where $T_* = \mathcal O(t^2/t_\perp)$ denotes the absolute energy scale in the honeycomb bilayer system and the factor $1/N^2$ accounts for the fact that self-energy effects are suppressed in the large-$N$ limit, cf.\ Eq.~\eqref{eq:slope-large-N}. In the quantum critical regime at $g \simeq g_\mathrm{c}$, there is a continuum of excitations and the specific heat $C_V$, for instance, will scale as
\begin{align}
	C_V \propto T^{d/z} \simeq 
	\begin{cases}
		T  &\text{for } T \gtrsim \frac{T_*}{N^2} \exp\!\left({-\frac{4\pi}{g_\mathrm{c} N}}\right), \\
		T^2 &\text{for } T \lesssim \frac{T_*}{N^2} \exp\!\left({-\frac{4\pi}{g_\mathrm{c} N}}\right).
	\end{cases}
\end{align}
At stronger couplings $g > g_\mathrm{c}$, there will be a finite-temperature phase transition towards an ordered state, assuming that the latter does not break a continuous symmetry. This is, for instance, the case for the inversion-symmetry-broken, charge density wave, or quantum anomalous Hall states discussed earlier. The critical temperature scales as $T_\mathrm{c} \propto (g - g_\mathrm{c})^{\nu z}$ with $z = 1$ and $\nu$ as in Eq.~\eqref{eq:nu-GNI} near the $(2+1)$D Gross-Neveu-Ising quantum critical point. The classical critical regime in the vicinity of the finite-temperature transition in this case is then described by the classical 2D Ising universality class, e.g., $\nu = 1$ and $\eta_\phi = 1/4$. It shrinks upon approaching $g \to g_\mathrm{c}$ from above.
Note that in the case of continuous symmetry breaking in the ordered ground state, such as in the spin-$1/2$ Hubbard model on the honeycomb bilayer for large on-site repulsion, there will be no genuine finite-temperature transition as a consequence of the Mermin-Wagner theorem. Nevertheless, the finite-temperature crossovers depicted in Fig.~\ref{fig:finite-T} will persist.


\section{Conclusions and outlook}
\label{sec:concl}

We have presented a theoretical analysis of 2D Fermi systems with quadratic band touching on lattices with $C_3$ symmetry. 
A natural physical realization is given by the problem of interacting fermions on Bernal-stacked bilayer honeycomb lattices, such as in bilayer graphene.
We have derived an effective low-energy continuum field theory that accounts for the explicit breaking of the continuous rotational symmetry characteristic for tricoordinate lattices and have shown, within a consistent perturbative RG calculation, that density-density interactions at two loops drive a splitting of each QBT point into four Dirac cones.
In contrast to the QBT systems with full rotational symmetry, in the systems with $C_3$ symmetry only, the semimetallic state is stable within a finite range of interactions $0 < g < g_\mathrm{c}$. At the critical coupling $g_\mathrm{c}$, the system undergoes a continuous quantum phase transition that has no classical analogue due to the presence of gapless fermion degrees of freedom at criticality.
This result is in agreement with previous quantum Monte Carlo~\cite{Pujarimain} and random phase approximation studies~\cite{honerkamp17b}.

The RG flow demonstrates that the quantum critical behavior near $g_\mathrm{c}$ is described by the celebrated Gross-Neveu-Ising (Gross-Neveu-Heisenberg) universality class for the case of Ising (Heisenberg) symmetry breaking, and we have given estimates for the universal critical exponents by employing known large-$N$ calculations~\cite{vasiliev93, gracey94, janssen14, gracey18}.
Our RG results have also uncovered the complex phenomenology at finite temperature, revealing crossovers between QBT, Dirac, and quantum critical regimes.
Furthermore, at small positive trigonal warping, $0 < t_\text{w} \ll t^2/t_\perp$, we have predicted an interesting sequence of three Gross-Neveu quantum phase transitions as a function of the short-range interaction.
All these predictions are directly testable using current numerical setups~\cite{lang12, Pujarimain}.

Bernal-stacked bilayer graphene exhibits an ordered ground state below $T_\mathrm{c} \approx 5\,\mathrm{K}$ with a zero-temperature gap $\Delta(0) \sim 3\,\text{meV}$~\cite{bao12}. The general scaling argument suggests $T_\mathrm{c} \sim T_* \exp(-\frac{4\pi}{g N})$, with the effective energy scale $T_*$, which may be estimated from the coefficient of the quadratic term in the dispersion [Eq.~\eqref{eq:dispQBT}] as $k_\mathrm{B} T_* \sim t^2/t_\perp \sim 20\,\mathrm{eV}$~\cite{zhang08}. From this, we estimate $g \sim 0.6$, which appears to be only slightly larger than our result for the critical coupling $g_\mathrm{c} \approx 0.4$ (see Sec.~\ref{subsec:phaseportrait}). This suggest that Bernal-stacked bilayer graphene may be not too far from the Gross-Neveu quantum critical point and that vestiges of the quantum critical scaling may be observable in a regime above the transition temperature, $T \gtrsim T_\mathrm{c}$. 
This applies, for instance, to transport properties such as the Hall coefficient $R_\mathrm{H}$, which in the disordered phase scales as $R_\mathrm{H}(T) \propto T^{-2/z}$ for $T > T_\mathrm{c}$. The dynamic exponent $z$ should then exhibit a crossover from $z \simeq 2$ for $T \gg T_\mathrm{c}$ to $z \simeq 1$ for $T \gtrsim T_\mathrm{c}$. Below the transition temperature, $T < T_\mathrm{c}$, the Hall coefficient will show an exponential behavior, $R_\mathrm{H}(T) \propto \exp[\frac{\Delta(T)}{k_\mathrm{B} T}]$.

Setups that allow one to tune the interaction strength experimentally should be able to reveal the quantum critical regime directly. It would be interesting to investigate this scenario, e.g., using cold atoms in an optical lattice~\cite{sun11}.

A worthwhile theoretical issue that we have neglected here, but may be relevant for bilayer graphene, is the effect of the long-range tail of the Coulomb repulsion. In the QBT limit, with vanishing trigonal warping, the density of states is finite and a long-range interaction is expected to be screened at low energy. When the QBT points split into Dirac cones due to the self-energy corrections, by contrast, screening is effectively suppressed. This might lead to a nontrivial interplay between the long-range and short-range components of the Coulomb interaction, potentially with similarities to the intriguing higher-dimensional case~\cite{moon13, herbut14, savary14, janssenherbutQBT, janssen16, janssen17, boettcher17}. %
It may also be useful to study the self-energy effects in the context of the competing-order problem occurring in realistic models for Bernal-stacked bilayer graphene~\cite{Cvetkovic}. To this end, one would need to extend the present single-channel analysis by employing a suitable Fierz-complete basis of four-fermion interactions~\cite{herbut09a, gies10, vafekbig} and studying the resulting interplay between these channels. This could lead to even richer physics at low and intermediate temperatures. %

\sr{\lj{Throughout this work, we have assumed particle-hole symmetry.} In real bilayer graphene, this \lj{will} be broken due to the presence of longer-ranged hopping terms. In that case, \lj{the Dirac cones generated dynamically from self-energy effects will form electron and hole pockets}. This \lj{might lead to further instabilites at the lowest temperatures and potentially new universality classes beyond the relativistic Gross-Neveu-Yukawa family.}}


\begin{acknowledgments}
We thank C.\ Honerkamp and M.\ M.\ Scherer for discussions and I.\ F.\ Herbut for valuable comments on the manuscript. %
This research was supported by the DFG through SFB 1143 (projects A02 and A04) and GRK 1621.
\end{acknowledgments}


\appendix
\begin{widetext}
\section{Position space propagator}
\label{app:loops}

We consider here the position space propagator $G_0(x)$, since it is the elementary building block as far as the evaluation of Feynman diagrams in real space is concerned. It satisfies the relation $\left[\partial_\tau + \mathcal{H}_0(-\rmi \nabla)\right] G_0(\tau,\vec{x}) = \delta^{(3)}(\tau,\vec{x})$. Translational invariance behooves us to solve it in Fourier space, to wit:
\begin{align}
  G_0(\tau,\vec{x}) = \int\frac{\rmd \omega\,\rmd^2\vec{p}}{(2\pi)^3}\,\rme^{\rmi \left(\omega \tau + \vec{p} \cdot \vec{x}\right)} \tilde{G}_0(\omega,\vec{p}),
\end{align}
with $\tilde{G}_0(\omega,\vec{p}) = \left[\rmi \omega + \mathcal{H}_0(\vec{p})\right]^{-1}$. The basic strategy now is to perform the Fourier integral in cylindrical coordinates
\[\vec{p} = (\rho \cos\varphi, \rho\sin\varphi); \quad \vec{x} = (r\cos\vartheta,r\sin\vartheta); \quad \vec{p}\cdot \vec{x} = r\rho \cos (\varphi - \vartheta).\]
For reasons of analytical tractability, we expanded the expression as a multilinear form in powers of rotational symmetry breaking ($f_1,f_3$ for the QBT theory and $f_2$ for the Dirac theory), keeping up to second order corrections, since that is the order to which we intend to evaluate all Feynman diagrams subsequently. Let us first consider the QBT limit $|f_1/f_2| \ll 1$. We parameterize the expanded propagator as
\begin{align}
\tilde{G}_0(\omega,\vec{p}) &= \sum_{\mu n m } 
\sigma^\mu \otimes \left(\sigma^3\right)^{\! n}
\left[\tilde{P}_{\mu nm}^{\kern.1em\text{c}}(\omega,\rho) \cos(m\varphi) + \tilde{P}_{\mu nm}^{\kern.1em\text{s}}(\omega,\rho) \sin(m\varphi)\right]\frac{\left(f_1 + \rho^2 f_3 \right)^{n}}{\left(\omega^2 + \rho^4\right)^{1+n}}
\end{align}
where $\mu \in \{0,1,2\}$, $k,m,n \in \mathbb{N}_{\geqslant 0}$, and $(\sigma^\mu) = (\mathds 1_2, \sigma_x, \sigma_y)$.
We have also set $f_2 = 1$ in the present QBT limit for convenience. We wish to expand to second order of rotational symmetry breaking, i.e., $n+m \leqslant 2$. The nonvanishing terms in $\tilde{G}_0(\omega,\vec{p})$ are then found to be:
\begin{gather}
\tilde{P}_{000}^{\kern.1em\text{c}}(\omega,\rho) = -\rmi \omega,\qquad
\tilde{P}_{102}^{\kern.1em\text{c}}(\omega,\rho) = \rho^2 = \tilde{P}_{102}^{\kern.1em\text{s}}(\omega,\rho),\qquad
\tilde{P}_{013}^{\kern.1em\text{c}}(\omega,\rho) = 2\rmi \omega \rho^3, \qquad
\tilde{P}_{111}^{\kern.1em\text{c}}(\omega,\rho) = \omega^2\rho = -\tilde{P}_{111}^{\kern.1em\text{s}}(\omega,\rho), \nonumber \\
\tilde{P}_{115}^{\kern.1em\text{c}}(\omega,\rho) = -\rho^5 = \tilde{P}_{115}^{\kern.1em\text{s}}(\omega,\rho),\qquad
\tilde{P}_{020}^{\kern.1em\text{c}}(\omega,\rho) = \rmi \left(\omega^3 \rho^2 - \omega \rho^6\right),\qquad
\tilde{P}_{026}^{\kern.1em\text{c}}(\omega,\rho) = -2\rmi \omega \rho^6, \nonumber \\
\tilde{P}_{122}^{\kern.1em\text{c}}(\omega,\rho) = -2\omega^2\rho^4 = \tilde{P}_{122}^{\kern.1em\text{s}}(\omega,\rho), \qquad
\tilde{P}_{124}^{\kern.1em\text{c}}(\omega,\rho) = -\omega^2\rho^4 = -\tilde{P}_{124}^{\kern.1em\text{c}}(\omega,\rho), \qquad
\tilde{P}_{128}^{\kern.1em\text{c}}(\omega,\rho) = \rho^8 = \tilde{P}_{128}^{\kern.1em\text{s}}(\omega,\rho).
\end{gather}
In the opposite Dirac limit, a similar expansion can be found in powers of $f_2$ (now setting $f_1 = 1$). The momentum space propagator is then parametrized as
\begin{align}
\tilde{G}_0(\omega,\vec{p}) &= \sum_{\mu n m} 
\sigma^\mu \otimes \left(\sigma^3\right)^{\! 1+n}
\left[\tilde{Q}_{\mu nm}^{\kern.05em\text{c}}(\omega,\rho) \cos(m\varphi) + \tilde{Q}_{\mu nm}^{\kern.05em\text{s}}(\omega,\rho) \sin(m\varphi)\right]\!
\frac{f_2 ^n}{\left(\omega^2 + \rho^2\right)^{1+n}}
\end{align}
where the $\tilde{Q}_{\mu nm}^{\kern.1em\text{c,s}}(\omega,\rho)$ again are bivariate polynomials, the nonvanishing ones among which are given by
\begin{gather}
\tilde{Q}^{\kern.05em\text{c}}_{000}(\omega,\rho) = -\rmi \omega, \qquad
\tilde{Q}^{\kern.05em\text{c}}_{101}(\omega,\rho) = \rho = -\tilde{Q}^{\kern.05em\text{s}}_{201}(\omega,\rho), \qquad
\tilde{Q}^{\kern.05em\text{c}}_{013}(\omega,\rho) = \omega \rho^3, \nonumber \\
\tilde{Q}^{\kern.05em\text{c}}_{112}(\omega,\rho) = \omega^2 \rho^2 = \tilde{Q}^{\kern.05em\text{s}}_{212}(\omega,\rho), \qquad
\tilde{Q}^{\kern.05em\text{c}}_{114}(\omega,\rho) = -\rho^4 = -\tilde{Q}^{\kern.05em\text{s}}_{212}(\omega,\rho), \qquad
\tilde{Q}^{\kern.05em\text{c}}_{020}(\omega,\rho) = \rmi\left(\omega^3 \rho^4 - \omega \rho^6\right), \nonumber \\
\tilde{Q}^{\kern.05em\text{c}}_{026}(\omega,\rho) = -2\rmi \omega \rho^6, \qquad
\tilde{Q}^{\kern.05em\text{c}}_{121}(\omega,\rho) = -2\omega^2\rho^5 = -\tilde{Q}^{\kern.05em\text{s}}_{221}(\omega,\rho), \nonumber \\
\tilde{Q}^{\kern.05em\text{c}}_{125}(\omega,\rho) = -\omega^2 \rho^5 = \tilde{Q}^{\kern.05em\text{s}}_{225}(\omega,\rho), \qquad
\tilde{Q}^{\kern.05em\text{c}}_{127}(\omega,\rho) = \rho^7 = -\tilde{Q}^{\kern.05em\text{s}}_{227}(\omega,\rho).
\end{gather}

In both cases, the Fourier integral with respect to $\omega$ is elementary. For the subsequent integral over $\varphi$, we employ the Jacobi-Anger expansion \cite{abrasteg} in the form
\begin{align}
\rme^{\rmi k\rho \cos(\varphi - \vartheta)} = {J}_0(k\rho) + 2\sum_{m=0}^\infty \rmi^m {J}_m(k\rho) \left[\cos(m\varphi)\cos(m\vartheta) + \sin(m\varphi)\sin(m\vartheta)\right],
\end{align}
where ${J}_m$ is the Bessel function of the first kind and order $m$, followed by the orthogonality relations of $\cos$ and $\sin$ in $L^2([0,2\pi])$. The remaining integral over $\rho$ turns out to be in fact expressible in terms of elementary functions, whence one obtains explicit expressions for the tree-level propagator in position space. We abstain from quoting them here in their full splendor due to their extraordinary length, and because they are not particularly enlightening.


\section{Pad\'e coefficients}
\label{app:pade}

Let us write the coefficient of $m$-th order in $(f_1/f_2)$ in the $\beta$ function of a quantity $X$ with $X \in \{g, (f_1/f_2), (f_3/f_2)\}$ as $\beta_{X}^{(\pm,m)}$ defined by
\begin{align}
\beta(X) &\simeq 
\begin{cases}
	\sum_{m \geqslant 0} \beta_{X}^{(+,m)}(f_1/f_2)^m & \text{for } f_1/f_2 \to 0, \\
	\sum_{m \geqslant 0} \beta_{X}^{(-,m)}(f_1/f_2)^{-m} & \text{for } f_1/f_2 \to \infty.
\end{cases}
\end{align}
Eqs.~\eqref{eq:betagQBT}--\eqref{eq:betafm1QBT} allow to read off $\beta_X^{(+,m)}$, while the dual coefficients $\beta_X^{(-,m)}$ can be read off from Eqs. \eqref{eq:betagDir}--\eqref{eq:betafDir}.

The Pad\'e coefficients defined in Eqs.~\eqref{eq:pade-1}--\eqref{eq:pade-3} are then given by
\begin{gather}
%
  a_0 = \beta_{f_1/f_2}^{(+,0)}, \qquad
  a_1 = \beta_{f_1/f_2}^{(+,1)}+\frac{\left(\beta_{f_1/f_2}^{(+,0)}\right)^{\!2}}{\left(\beta_{f_1/f_2}^{(-,1)}\right)^{\!2} - \beta_{f_1/f_2}^{(+,1)}\beta_{f_1/f_2}^{(-,1)}}, \qquad
  a_2 = \frac{\beta_{f_1/f_2}^{(+,0)}}{\beta_{f_1/f_2}^{(-,1)}-\beta_{f_1/f_2}^{(+,1)}}, \nonumber \\
  b_1 =  \frac{\beta_{f_1/f_2}^{(+,0)}}{\left(\beta_{f_1/f_2}^{(-,1)}\right)^{\!2} - \beta_{f_1/f_2}^{(+,1)}\beta_{f_1/f_2}^{(-,1)}}, \qquad
  b_2 = \frac{1}{\beta_{f_1/f_2}^{(-,1)}}, \qquad
  c_0 = \beta_{f_3/f_2}^{(+,0)}, \qquad
  c_1 = \beta_{f_3/f_2}^{(+,1)}, \qquad
  c_2 = \beta_{f_3/f_2}^{(-,0)}, \nonumber \\
  d_0 = \beta_{g}^{(+,0)}, \qquad
  d_1 = \beta_{g}^{(-,1)}, \qquad
  d_2 = \beta_{g}^{(+,2)} + \beta_{g}^{(+,0)}\frac{\beta_{g}^{(-,2)} - \beta_{g}^{(+,2)}}{\beta_{g}^{(+,0)} - \beta_{g}^{(-,0)}}, \qquad
  d_4 = \beta_{g}^{(-,0)}, \qquad
  e_2 = \frac{\beta_{g}^{(-,2)} - \beta_{g}^{(+,2)}}{\beta_{g}^{(+,0)} - \beta_{g}^{(-,0)}}.
%
\end{gather}

\end{widetext}




\begin{thebibliography}{99}

\bibitem{castrorev} A. H. Castro Neto, F. Guinea, N. M. R. Peres, K. S. Novoselov, and A. K. Geim,
Rev. Mod. Phys. {\bf 81}, 109 (2009).

\bibitem{herbut06} I. F. Herbut, 
Phys. Rev. Lett. {\bf 97}, 146401 (2006).

\bibitem{herbut09} I. F. Herbut, V. Juri\v{c}i\'{c}, and O. Vafek,
Phys. Rev. B {\bf 80}, 075432 (2009).

\bibitem{assaad13} F. F. Assaad and I. F. Herbut,
Phys. Rev. X {\bf 3}, 031010 (2013).

\bibitem{janssen14} L. Janssen and I. F. Herbut,
Phys. Rev. B {\bf 89}, 205403 (2014).

\bibitem{otsuka16} Y. Otsuka, S. Yunoki, and S. Sorella,
Phys. Rev. X {\bf 6}, 011029 (2016).

\bibitem{ref6} K. Sun, H. Yao, E. Fradkin, and S. A. Kivelson,
Phys. Rev. Lett. {\bf 103}, 046811 (2009).

\bibitem{herbut14} I. F. Herbut and L. Janssen,
Phys. Rev. Lett. {\bf 113}, 106401 (2014).

\bibitem{McCannFalko} E. McCann, and V. I. Fal'ko,
Phys. Rev. Lett. {\bf 96} 086805 (2006).

\bibitem{Cvetkovic} V. Cvetkovic, R. E. Throckmorton, and O. Vafek, 
Phys. Rev. B {\bf 86}, 075467 (2012).

\bibitem{rozhkov16} A. V. Rozhkov, A. O. Sboychakov, A. L. Rakhmanov, and F. Nori,
Phys. Rep. {\bf 648}, 1 (2016).

\bibitem{feldman09} B. E. Feldman, J. Martin, and A. Yacoby,
Nat. Phys. {\bf 5}, 889 (2009).

\bibitem{martin10} J. Martin, B. E. Feldman, R. T. Weitz, M. T. Allen, and A. Yacoby,
Phys. Rev. Lett. {\bf 105}, 256806 (2010).

\bibitem{weitz10} R. T. Weitz, M. T. Allen, B. E. Feldman, J. Martin, and A. Yacoby, 
Science {\bf 330}, 812 (2010).

\bibitem{velasco12} J. Velasco Jr {\it et al.}, %
Nat. Nanotechnol. {\bf 7}, 156 (2012).

\bibitem{freitag12} F. Freitag, J. Trbovic, M. Weiss, and C. Sch\"{o}nenberger, 
Phys. Rev. Lett. {\bf 108}, 076602 (2012).

\bibitem{bao12} W. Bao {\it et al.}, 
Proc. Natl. Acad. Sci. U.S.A. {\bf 109}, 10802 (2012).

\bibitem{veligura12} A. Veligura, H. J. van Elferen, N. Tombros, J. C. Maan, U. Zeitler, and B. J. van Wees, 
Phys. Rev. B {\bf 85}, 155412 (2012).

\bibitem{mayorov11} A. S. Mayorov {\it et al.}, 
Science {\bf 333}, 860 (2011).

\bibitem{Pujarimain} S. Pujari, T. C. Lang, G. Murthy, and R. K. Kaul,
Phys. Rev. Lett. {\bf 117}, 086404 (2016).

\bibitem{honerkamp17b} C. Honerkamp,
Phys. Rev. B {\bf 96}, 245134 (2017).

\bibitem{zhang10} F. Zhang, H. Min, M. Polini, and A. H. MacDonald,
Phys. Rev. B {\bf 81}, 041402 (2010).

\bibitem{vafekyang} O. Vafek and K. Yang,
Phys. Rev. B {\bf 81}, 041401 (2010).

\bibitem{vafekbig} O. Vafek,
Phys. Rev. B {\bf 82}, 205106 (2010).

\bibitem{uebelacker11} S. Uebelacker and C. Honerkamp,
Phys. Rev. B {\bf 84}, 205122 (2011).

\bibitem{lang12} T. C. Lang, Z. Y. Meng, M. M. Scherer, S. Uebelacker, F. F. Assaad, A. Muramatsu, C. Honerkamp, and S. Wessel, 
Phys. Rev. Lett. {\bf 109}, 126402 (2012).

\bibitem{scherer12} M. M. Scherer, S. Uebelacker, and C. Honerkamp,
Phys. Rev. B {\bf 85}, 235408 (2012).

\bibitem{song12} K. W. Song, Y.-C. Liang, and S. Haas,
Phys. Rev. B {\bf 86}, 205418 (2012).

\bibitem{sinnerziegler} A. Sinner, and K. Ziegler,
Phys. Rev. B {\bf 82}, 165453 (2010).

\bibitem{McCannKoshino} E. McCann, and M. Koshino,
Rep. Prog. Phys. {\bf 76}, 056503 (2013).

\bibitem{janssenherbutQBT} L. Janssen and I. F. Herbut,
Phys. Rev. B {\bf 92}, 045117 (2015).

\bibitem{herbut09a} I. F. Herbut, V. Juri\v{c}i\'{c}, and B. Roy,
Phys. Rev. B {\bf 79}, 085116 (2009).

\bibitem{gies10} H. Gies and L. Janssen,
Phys. Rev. D {\bf 82}, 085018 (2010).

\bibitem{hands93} S. Hands, A. Koci\'{c}, and J. B. Kogut,
Ann. Phys. (N. Y.) {\bf 224}, 29 (1993).

\bibitem{vasiliev93} A. Vasiliev, S. E. Derkachov, N. Kivel, and A. Stepanenko,
Theor. Math. Phys. {\bf 94}, 127 (1993).

\bibitem{gracey94} J. A. Gracey, 
Int. J. Mod. Phys. A {\bf 09}, 727 (1994).

\bibitem{vojta00a} M. Vojta, Y. Zhang, and S. Sachdev,
Phys. Rev. B {\bf 62}, 6721 (2000).

\bibitem{vojta00b} M. Vojta, Y. Zhang, and S. Sachdev,
Phys. Rev. Lett. {\bf 85}, 4940 (2000).

\bibitem{braun11} J. Braun, H. Gies, and D. D. Scherer,
Phys. Rev. D {\bf 83}, 085012 (2011).

\bibitem{gracey16} J. A. Gracey, T. Luthe, and Y. Schr\"{o}der,
Phys. Rev. D {\bf 94}, 125028 (2016).

\bibitem{mihaila17} L. N. Mihaila, N. Zerf, B. Ihrig, I. F. Herbut, and M. M. Scherer,
Phys. Rev. B {\bf 96}, 165133 (2017).

\bibitem{zerf17} N. Zerf, L. N. Mihaila, P. Marquard, I. F. Herbut, and M. M. Scherer,
Phys. Rev. D {\bf 96}, 096010 (2017).

\bibitem{iliesiu18} L. Iliesiu, F. Kos, D. Poland, S. S. Pufu, and D. Simmons-Duffin,
J. High Energy Phys. 01 (2018) 036.

\bibitem{ihrig18} B. Ihrig, L. N. Mihaila, and M. M. Scherer,
Phys. Rev. B {\bf 98}, 125109 (2018).

\bibitem{sunrisexspace} S. Groote, J. G. K\"orner, and A. A. Pivovarov,
Nucl. Phys. B {\bf 542}, 515 (1999).

\bibitem{gehring15} F. Gehring, H. Gies, and L. Janssen,
Phys. Rev. D {\bf 92}, 085046 (2015).

\bibitem{zhu16} W. Zhu, S.-S. Gong, T.-S. Zeng, L. Fu, and D. N. Sheng,
Phys. Rev. Lett. {\bf 117}, 096402 (2016).

\bibitem{zeng18} T.-S. Zeng, W. Zhu, and D. Sheng,
npj Quantum Mater. {\bf 3}, 49 (2018).

\bibitem{sur18} S. Sur, S.-S. Gong, K. Yang, and O. Vafek,
Phys. Rev. B {\bf 98}, 125144 (2018).

\bibitem{herbutbook} I. Herbut, 
{\it A Modern Approach to Critical Phenomena}
(Cambridge University Press, Cambridge, U.K., 2007).

\bibitem{toldin15} F. Parisen Toldin, M. Hohenadler, F. F. Assaad, and I. F. Herbut,
Phys. Rev. B {\bf 91}, 165108 (2015).

\bibitem{knorr18} B. Knorr, 
Phys. Rev. B {\bf 97}, 075129 (2018).

\bibitem{gracey18} J. A. Gracey,
Phys. Rev. D {\bf 97}, 105009 (2018).

\bibitem{zhang08} L. M. Zhang, Z. Q. Li, D. N. Basov, M. M. Fogler, Z. Hao, and M. C. Martin,
Phys. Rev. B {\bf 78}, 235408 (2008).

\bibitem{sun11} K. Sun, W. V. Liu, A. Hemmerich, and S. Das Sarma,
Nat. Phys. {\bf 8}, 67 (2012).

\bibitem{moon13} E.-G. Moon, C. Xu, Y. B. Kim, and L. Balents,
Phys. Rev. Lett. {\bf 111}, 206401 (2013).

\bibitem{savary14} L. Savary, E.-G. Moon, and L. Balents,
Phys. Rev. X {\bf 4}, 041027 (2014).

\bibitem{janssen16} L. Janssen and I. F. Herbut,
Phys. Rev. B {\bf 93}, 165109 (2016).

\bibitem{janssen17} L. Janssen and I. F. Herbut,
Phys. Rev. B {\bf 95}, 075101 (2017).

\bibitem{boettcher17} I. Boettcher and I. F. Herbut,
Phys. Rev. B {\bf 95}, 075149 (2017).

\bibitem{abrasteg} M. Abramowitz and I. A. Stegun (eds), {\it Handbook of Mathematical Functions with Formulas, Graphs, and Mathematical Tables} (Dover Publications, New York, U.S.A., 1972).


\end{thebibliography}
\end{document}